\documentclass[showpacs,preprintnumbers,amsmath,amssymb,12pt,floatfix]{revtex4}
\usepackage[pdftex]{graphicx}
\usepackage[usenames, dvipsnames]{color}
\usepackage{amssymb}
\usepackage{bm}
\usepackage{graphicx}
\usepackage{revsymb}
\usepackage{amsmath}
\usepackage{dcolumn}
 \usepackage{ulem}

\usepackage{color} 

\def\be{\begin{equation}}
\def\ee{\end{equation}}
\def\bea{\begin{eqnarray}}
\def\eea{\end{eqnarray}}

\begin{document}

\title{Symmetry Energy from Two-Nucleon Separation Energies of Pb and Ca Isotopes}
\author{Myeong-Hwan Mun and Myung-Ki Cheoun \footnote{cheoun@ssu.ac.kr (Corresponding Author)}}
\affiliation{Department of Physics and Origin of Matter and Evolution of Galaxies (OMEG) Institute, Soongsil University, Seoul 06978, Korea}

\author{Eunja Ha} 
\affiliation{Department of Physics and Research Institute for Natural Science, Hanyang University, Seoul 04763, Korea}


\author{H. Sagawa} 
\affiliation{RIKEN, Nishina Center for Accelerator-Based Science, Wako 351-0198, Japan and Center for Mathematics and Physics, University of Aizu, Aizu-Wakamatsu, Fukushima 965-8560, Japan}

\author{Gianluca Col\`o} 
\affiliation{Dipartimento di Fisica, Università degli Studi, and INFN sezione di Milano, via Celoria 16, 20133 Milano, Italy}

\date{\today}

\begin{abstract}
We investigate the symmetry energy in relation with the two-proton and  two-neutron separation energies using different  nuclear mass  data. For this aim,  we exploit the deformed relativistic Hartree-Bogoliubov theory in the continuum (DRHBc), FRDM2012 and AME2020 data. 
First, we study the two-proton and two-neutron separation energies in Pb and Ca isotopes by subtracting the contribution of Coulomb energy.
They show a strong correlation with neutron number as well as with the neutron skin thickness. By taking the relative difference of both separation energies, we derive the symmetry energy from Ca and Pb isotopes. Since the nuclear surface contributes to the symmetry energy, we  deduce the volume symmetry energy by subtracting the surface contribution using several mass models. The obtained symmetry energy coefficient, $a_{sym}$, is 20.0 $\sim $ 22.7 MeV for Pb isotopes and 18.7 $\sim$ 19.3 MeV for Ca isotopes from the DRHBc mass table data, while the results from other mass tables are 19.6 $\sim$ 22.1 (20.7 $\sim$ 22.3) MeV for Pb isotopes and 18.9 $\sim$ 19.0 (19.6 $\sim$ 19.7) MeV for Ca isotopes from AME2020 (FRDM2012) data. The volume contribution to the asymmetry coefficient, $a_{sym}^v$, which depends on the ratio of the surface to the volume energy coefficients, $a_s / a_v$, is also provided for each mass model. Since the ratio $a_s / a_v$ is neither determined by nuclear theory, nor by experimental data, we have investigated $a_{sym}^v$ by using the ratio $a_s / a_v$ as a free parameter, and have obtained $a_{sym}^v = $ 27.0 MeV, almost irrespective of nuclear model and isotopic chain, with the ratio  $a_s / a_v$ constrained as $a_s / a_v = 1.10 \sim 1.13$.               
\end{abstract}

\maketitle

\section{Introduction}
The neutron skin thickness (NST) of neutron-rich nuclei has been one of the most  important topics in nuclear physics because it can provide critical information on the symmetry energy in finite nuclei,  as well as on the structure of neutron stars.  There are several experimental  data for the NST, such as parity-violating electron scattering, PREX I, PREX II and CREX, as well as data on the dipole polarizability obtained by proton elastic scattering, and anti-proton scattering \cite{Adhikari2021,Zenihiro_arXiv,Zenihiro2010,Starodubsky1994,Ray1978}. From the viewpoint of the symmetry energy, the NST is a significant observable because it is proportional to the slope parameter of the equation of state (EoS) of nuclear matter.

In this work, we focus on the proton and neutron separation energy because the proton separation energy implies the propagation of proton in the NST region, and consequently it is affected by the NST, while the neutron separation energy becomes smaller with the increase of the neutron number due to the more dilute neutron density in the NST. Specifically, we utilize two-neutron and two-proton separation energies in even-even Pb and Ca isotopes because one-proton and one-neutron separation energies need to consider the additional pairing energy of even-odd nuclei, and the Pauli blocking. In addition, since the Coulomb energy is crucial to determine the proton separation energy, we need to subtract the contribution from those separation energies to discuss the symmetry energy of finite nuclei. Furthermore, for the symmetry energy in nuclear matter, we derive the volume symmetry energy coefficient subtracted by the surface contribution.   

By using the mass table of the DRHBc model, which attained more neutron-rich and proton-rich nuclei than other nuclear mass models, we can deduce the two-nucleon separation energies in a wide mass region, including nuclei near the drip lines. Then, the difference has an explicit relation with the  asymmetry energy coefficient, $a_{sym}$, in the semi-empirical Bethe-Weizs\"acker liquid drop mass model. Based on the  Bethe-Weizs\"acker  mass formula, the total binding energy of a nucleus is written as
\begin{equation} \label{BW-mass}
BE(A, Z) = a_{v}A - a_{s}A^{2/3} - a_{c} {Z^{2} \over A^{1/3}} - a_{sym} I^{2}A, 
\end{equation}
for a nucleus with mass number $A$ and proton number $Z$.  In Eq. \eqref{BW-mass},  $a_v, a_s, a_c,$ and  $a_{sym}$  are the volume, surface, Coulomb and symmetry energy terms, respectively.    $I = (N - Z) / A $ represents the isospin asymmetry and the  pairing energy term is discarded since it does not play any role in our further considerations.
 Here, we use minus signs except for the volume term, by which the binding energy and all coefficients are positive, because the surface, the Coulomb, and the symmetry energy tend to unbind the nucleus. By subtracting the Coulomb energy, we define a modified binding energy  formula without the Coulomb contribution as 
\begin{equation}
BE^{*}(A, Z) = BE(A, Z) + a_{c} {Z^{2} \over A^{1/3}}  =  a_{v}A - a_{s}A^{2/3}  - a_{sym} I^{2}A .
\end{equation}
Then, the separation energies without the Coulomb energy term are given by
\begin{equation}
S_{2n}^{*} = [BE^{*}(A, Z) - BE^{*}(A-2, Z)] ~,~ S_{2p}^{*} = [BE^{*}(A, Z) - BE^{*}(A-2, Z-2)] .
\end{equation}
{The difference between the corrected $2p$ and $2n$ separation energies} {can be} expressed in term of the isospin asymmetry coefficient as
\begin{equation}
S_{2p}^{*} - S_{2n}^{*} = a_{sym} {8(N - Z) \over (A-2)} =  8a_{sym} I^{*},~~{\rm with}~~I^{*}={N - Z \over A-2}~.   
\label{eq:sseq1}
\end{equation}

For the application to nuclear matter, we divide the symmetry energy into the volume and surface part ($a_{sym}^{s}$, $a_{sym}^v$), using the following relation \cite{Lunn2003}: $a_{sym} I^2 A = ( a_{sym}^{v} A + a_{sym}^{s} A^{2/3})  I^{2}$. Then Eq. (\ref{eq:sseq1}) is reexpressed as
\begin{equation}
 S_{2p}^{*} - S_{2n}^{*}               
             =  8I^{*}(a_{sym}^{v} + a_{sym}^{s}(A-2)^{-1/3}) = 8 I^* a_{sym}^{v}(1 - {a_{s} \over a_{v}}(A-2)^{-1/3})  ~,       
\label{eq:sseq2}
\end{equation}
where {we define the ratio of the surface to the volume term as $a_{sym}^s / a_{sym}^v = - a_s/a_v$} \cite{Lunn2003}.

We take $a_{v}$ and $a_{s}$, whose values depends on the nuclear model, as summarized at Table II in Appendix D of Ref. \cite{Lunn2003}. We have performed calculations for Pb and Ca isotopes using the DRHBc mass model \cite{Kaiyuan2022}, the FRDM2012 mass model \cite{FRDM}, and the AME2020 data \cite{AME2020}. The results show an interesting {consequence} for the symmetry energy in finite nuclei.

This paper is organized as follows. A brief description of the DRHBc theory used in the present calculation is presented in Sec. II. Detailed results of Pb and Ca isotopes including the NST and {the nucleon separation energies} are provided in Sec. III. Finally, the summary and conclusion are given  in Sec. IV.

\section{FORMALISM}
In order to {see the consequences of the above discussion}, we need a well-refined nuclear model which has to incorporate the deformation, the pairing correlations and the continuum, through a microscopic approach, 
{so that it can account well for properties of the nuclear masses as a whole, by}
covering nuclei near the drip lines.
Another important ingredient is the relativistic description which has been initiated by the authors in Refs. \cite{Walecka1974,Boguta1977} with various meson-exchange models inside nuclei, and has enabled us to incorporate consistently the nucleonic spin degree of freedom.

Along this line,  the deformed relativistic Hartree-Bogoliubov theory in continuum (DRHBc) was developed for deformed halo nuclei in Refs. \cite{Zhou2010,Lulu2012}, and recently extended \cite{Kaiyuan2020} with point-coupling density functionals. This theory is proved to be capable {of providing a good} description of the nuclear masses with high predictive power \cite{Kaiyuan2021,Cong2021,Kaiyuan2022}, and {it has 
successfully been applied to some particular} nuclei \cite{Cong2019,Sun2018,Sun2020,In2021,Yang2021,SunPRC1,SunPRC2,Sun2021}. It has followed the previous relativistic continuum Hartree-Bogoliubov (RCHB) approach, calculated in coordinate space \cite{Meng1996,Meng1998} by explicitly including the deformation in a Dirac Woods-Saxon basis \cite{Zhou2003}. Here we note that the deformed cylindrical basis preserving axial symmetry could be an alternative to effectively treat the convergence of the total energy as argued in Ref. \cite{Robledo2012}, where the Gogny-type pairing force was exploited for neutron rich nuclei near drip lines and odd nuclei.

In this work, we focus on the two-nucleon separation energies and NST of the Pb and Ca isotopes within the DRHBc theory, which was succinctly summarized in Refs. \cite{Kaiyuan2020,Lulu2012}. The present calculations are carried out in the following relativistic Hartree-Bogoliubov theory with the density functional PC-PK1 \cite{Zhao2010},

\begin{equation} \label{eq:hfbeq}
\left( \begin{array}{cc} h_D - \lambda &
\Delta  \\
 - \Delta^{*} & - h_D^{*} + \lambda
  \end{array}\right)
\left( \begin{array}{c}
U_{k} \\ V_{k}  \end{array}\right)
 =
 E_{k}
\left( \begin{array}{c} U_{k} \\
V_{k} \end{array}\right),
\end{equation}
where $h_D, \lambda, E_k, (U_k,V_k)$ are the Dirac Hamiltonian, the Fermi energy and the quasiparticle energy and wave function, respectively.
The pairing potential $\Delta$ is given {as a function of} the pairing tensor $\kappa ({\bf r}, {\bf r}^{'})$ as follows
\begin{equation}
\Delta({\bf r}, {\bf r}^{'}) = V ({\bf r}, {\bf r}^{'}) \kappa ({\bf r}, {\bf r}^{'}), 
\end{equation}
with a density-dependent zero range force
\begin{equation}
V({\bf r}, {\bf r}^{'}) = {V_0 \over 2} ( 1 - P_{\sigma}) \delta ( {\bf r} - {\bf r}^{'}) ( 1 -  { \rho (\bf r) \over \rho_{sat}} )~.
\end{equation}

For the pairing strength, we use $V_0$ = -- 325.0 MeV fm${^3}$. The saturation density is adopted as $\rho_{sat}$ = 0.152 fm$^{-3}$, together with a pairing window 
$E_{cut}^{q.p}$ = 100 MeV. The energy cut-off $E_{cut}^+ =$ 300 MeV,  and the angular momentum cut-off $J_{max} = (23/2) \hbar $, are taken for the Dirac Woods-Saxon basis. The above numerical details are the same as those suggested in Refs. \cite{Kaiyuan2020,Kaiyuan2021} for the DRHBc mass table calculation. For the present calculation of the Pb (Ca) isotopes, the Legendre expansion truncation is chosen as ${\lambda}_{max}$ = 8 (6) \cite{Kaiyuan2020,Kaiyuan2021}.

Empirical pairing gaps of Pb isotopes were shown to be properly reproduced with the energy cut-off, the maximum angular momentum and the Legendre expansion truncation obtained from the convergence check of total energies, as shown in Fig. 5(b) in Ref. \cite{Kaiyuan2020}.

The present zero-range scheme for the pairing force is better than the simple constant gap approximation, but it has still the pairing window problem in the pairing tensor, as discussed in 
Refs. \cite{Dobaczewski1996,Tian2009}, because it needs an arbitrary energy cut-off parameter for neutron-rich nuclei. In spherical nuclei, the neutron pairing gaps are well reproduced by the pairing window defined by $E_{cut}^{q.p}$ = 100 MeV. But for deformed nuclei $A =$ 186 $\sim$ 198 (see Fig. 2 in Ref. \cite{Kim2022}) the results of $E_{cut}^{q.p}$ = 200 MeV are better than those by other windows. This implies that the convergence of total energies with the pairing windows has to be more carefully assessed in the case of deformed nuclei. We leave it as a future work. More elaborate approach for the pairing interaction, beyond the zero-range scheme, is that based on the Gogny-type finite-range pairing force \cite{Decharge1980}, using a separable approximation \cite{Niksic2014,Serra2002}. This kind of treatment of pairing interaction is also applied to the covariant density-functional-theory (DFT) adopted for studying neutron-rich nuclei in Refs. \cite{AfanasjevPRL,Afanasjev2PRC,Vretenar2005,Agbemava2014}.

\section{RESULTS AND DISCUSSION}

\subsection{Two-neutron and two-proton separation energies}

\begin{figure}
\centering
\includegraphics[width=0.95\linewidth]{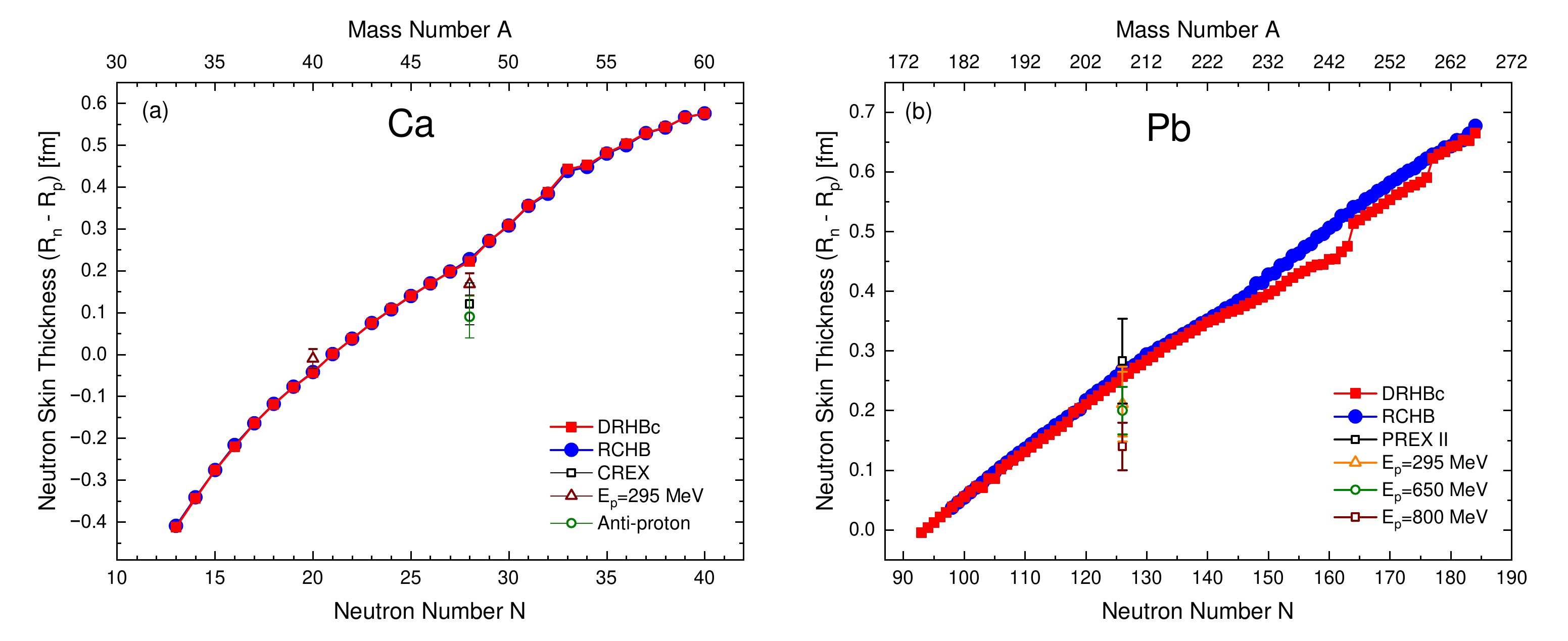}
\caption{(Color online) Evolution of NST for Ca (a) and Pb (b) isotopes. We obtain the NST for $^{208}$Pb as $R_n - R_p$ = 0.257 fm and the NST for $^{40}$Ca as $R_n - R_p = - 0.043$ fm in the DRHBc results. {Data from different experiments (CREX and PREX II, anti-proton scattering, and proton elastic scattering at different incident proton energies $E_p$) are also displayed, and detailed in Table \ref{tableNST}.}
In theoretical studies, DRHBc takes into account both the continuum and the deformation, while RCHB includes only the continuum effect.}
\label{fig1}
\end{figure}

Both the DRHBc and RCHB results show a clear correlation between the NST and the neutron number, as shown in Fig.~\ref{fig1}.
The small deviation of the RCHB results \cite{RCHB2018} in Pb isotopes, with respect to the DRHBc results, can be attributed to the fact that the RCHB model does not consider explicitly the deformation.

We obtain the NST of $^{40,48}$Ca equal to --0.043 fm and 0.223 fm, respectively, in the  DRHBc model. For $^{208}$Pb, the DRHBc value is 0.257 fm. 
We compare these results with various experimental findings both in Fig.~\ref{fig1} and in Table~\ref{tableNST}.
An interesting point is that 
$^{208}$Pb data from proton elastic scattering experiments 
show systematically smaller NST than the data from PREX II, as well as than the results of the present calculations.  On the other hand, such data are consistent with the dipole polarization experiments and 
also with non-relativistic Skyrme calculations.  Curiously, the NST from proton scattering data of $^{48}$Ca is larger than 
  that of CREX data, but smaller than the DRHBc results.
 While there might be model dependence related to the optical model adopted in the analysis of the proton elastic scattering data, the difficulty to reconcile PREX and CREX data is still an interesting and open problem.

\begin{table}[h]
	\caption{Comparison of the NST to the experimental data in the unit of fm. 
	The value of 
	$E_p$ for Ca is 295 MeV, and  there are different values, namely 295, 650 and 800 MeV for $^{208}$Pb \cite{Zenihiro2010,Starodubsky1994,Ray1978}. The anti-proton scattering data are from Ref. \cite{Trzc2001}.}
	\begin{tabular}{c|c|c|c|c}
		\hline
		\hline
		& DRHBc    & p-elastic scatt. data $(E_p)$ [MeV] & PREX II and CREX     & ${\bar p}$ data         \\ 
		\hline
		$^{40}$Ca  & --0.043 & -- 0.010$^{+0.022}_{-0.024}$  (295)        &~~~$-$~~ &  $-$              \\ 
		\hline
		$^{48}$Ca  & 0.223  & 0.168$^{+0.025}_{-0.028}$ (295)           & 0.121 $\pm$ 0.026(exp) $\pm$ 0.024(model) & 0.090 $\pm$ 0.050 \\ 
		\hline
		&          & 0.211$^{+0.054}_{-0.063}$  (295) &                     &               \\
		$^{208}$Pb & 0.257    & 0.20 $\pm$ 0.04  (650)           & 0.283 ± 0.071       & ~~~$-$~~      \\
		&          & 0.14 $\pm$ 0.04 (800)            &                     &               \\ 
		\hline
		\hline
	\end{tabular}
	\label{tableNST}
\end{table}

\begin{figure}
\centering
\includegraphics[width=0.95\linewidth]{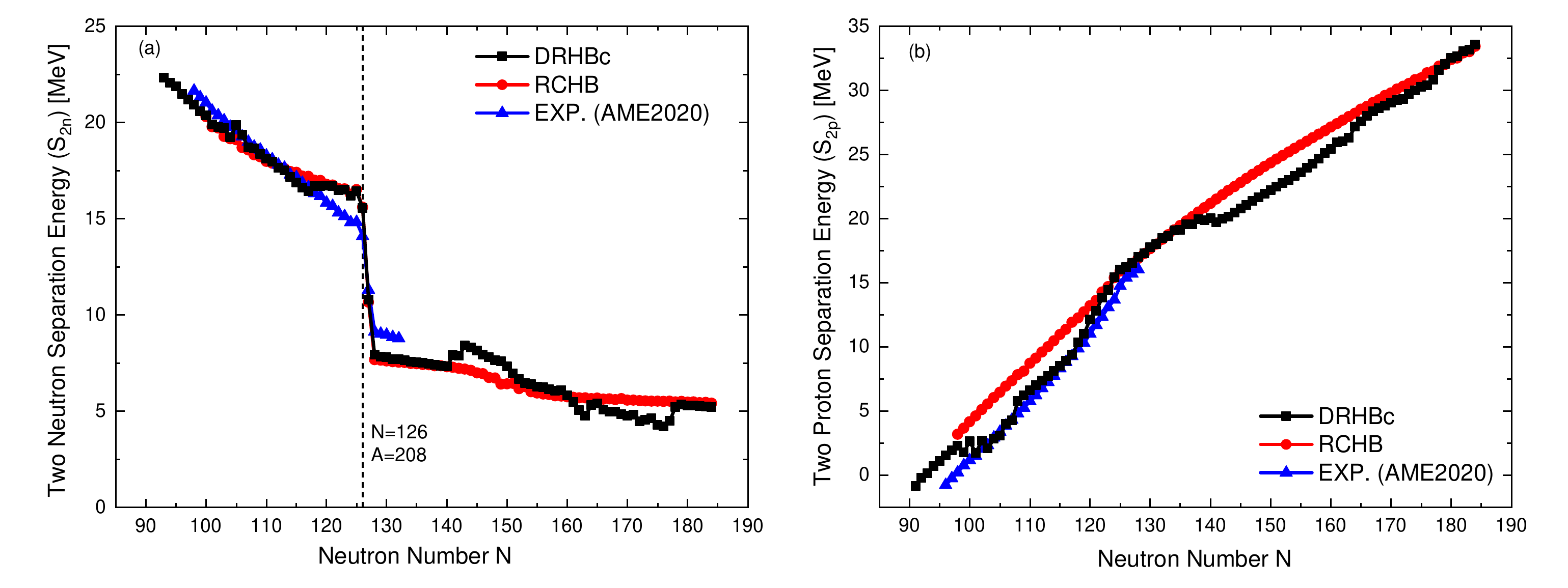}
\caption{(Color online) Evolution of two-nucleon separation energies, $S_{2n}$ and $S_{2p}$, for Pb isotopes, compared to the DRHBc data \cite{Kaiyuan2022}, RCHB data \cite{RCHB2018}, and AME2020 data \cite{AME2020}.} 
\label{fig2}
\end{figure}

Figure \ref{fig2} shows the two-neutron separation energy ($S_{2n}$) and two-proton separation energy ($S_{2p}$) in terms of neutron number. $S_{2n}$ decreases with the neutron number and shows an abrupt decrease at the neutron magic number $N = 126$. This is an indication of the magic shell structure. In order to extract the symmetry energy coefficient, we have to subtract   
  the shell corrections, which has been usually done by the Strutinsky method \cite{Stru1967}.    
Here we adopt a simple smooth continuation approach, by  subtracting   the Fermi energy difference of nuclei before and after the shell closure.

We also note an almost monotonic increase of $S_{2p}$ with the neutron number. With the increase of the NST by the increase of neutron number, {because of the strong proton-neutron interaction, the proton potential is more deepened,} and consequently $S_{2p}$ increases. This  can also be explained by the increase of Fermi energy difference between proton and neutron, $\epsilon_{p} - \epsilon_{n}$, as discussed in Ref. \cite{Ning2013}. Furthermore, the monotonic increase of $S_{2p}$ with the neutron number implies that the NST, as well as the slope parameter of the symmetry energy, are closely related to the behaviour of $S_{2p}$, which will be discussed in detail later on.        

\begin{figure}
\centering
\includegraphics[width=0.95\linewidth]{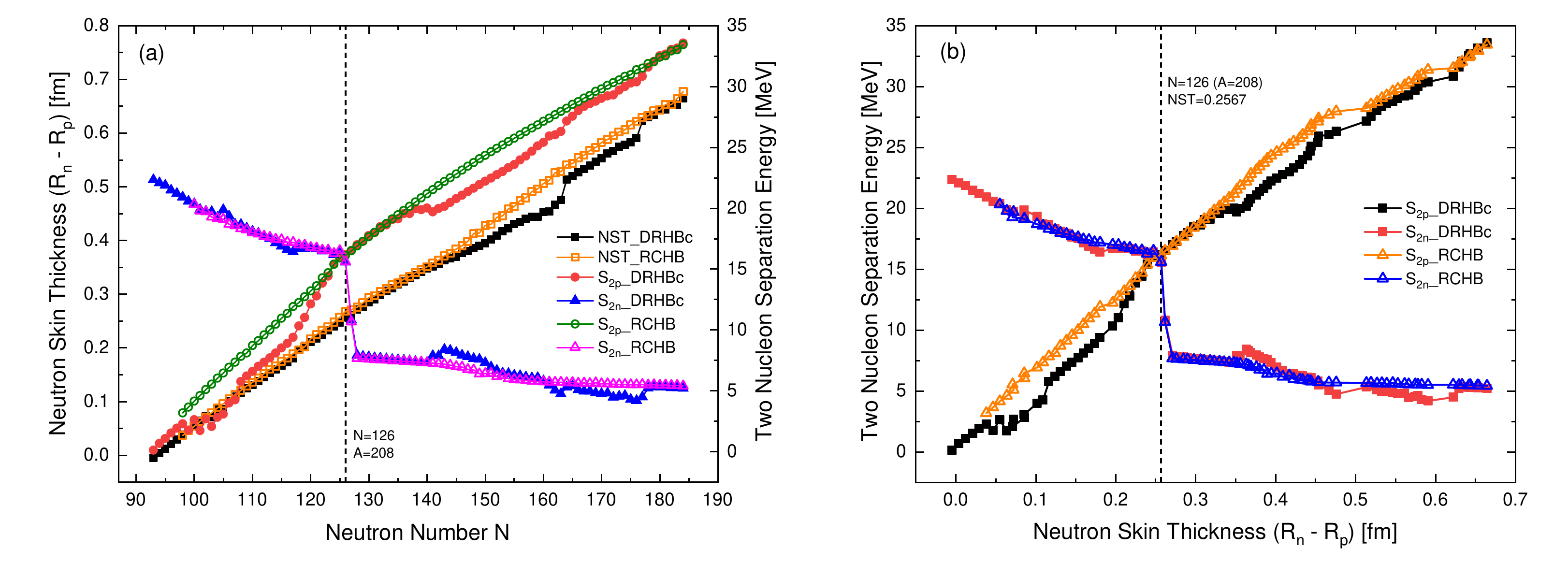}
\caption{(Color online) (a) Evolution of the NST, $S_{2p}$ and $S_{2n}$ with the neutron number (denoted by squares, circles and triangles, respectively); (b) evolution of $S_{2n}$ and $S_{2p}$ with the NST, for Pb isotopes, obtained by DRHBc \cite{Kaiyuan2022} and RCHB \cite{RCHB2018}.}
\label{fig3}
\end{figure}

Fig.~\ref{fig3} demonstrates the evolution of the NST with the neutron number, and that of $S_{2n}$ and $S_{2p}$ for Pb isotopes with the NST. All observables  show clear correlations with the neutron number. In particular, Fig. \ref{fig3}(a) shows different increase rates of the NST and $S_{2p}$. Interestingly, $S_{2p}$ shows steeper increase with the neutron number than the NST. Since the neutron number and the NST also have a correlation, we show the evolution of the $S_{2p}$ and $S_{2n}$ with the NST in Fig. \ref{fig3}(b). Other NST calculations by RCHB model \cite{Meng1996,Meng1998} show more straight correlations to the neutron number, and the separation energies also display similar pattern as the NST.  
The smooth correlations displayed by the RCHB results stem from the lack of deformation in these calculations.

Hereafter, we focus on the evolution of $S_{2n}$ and $S_{2p}$ with the neutron number. A detailed discussion of the relation to the NST will be presented in a forthcoming paper, with the additional analysis of the symmetry energy in nuclear matter.


\begin{table}[h]
\caption{{Formulas for the Coulomb energy corrections to Eq. (1) in Ref. \cite{Tian2014},  where  the differences of  experimental
binding energies between  88 pairs of mirror nuclei (with the same
mass number A, but with neutrons and protons interchanged) are adopted to fix the parameters $a_c$ and $b$ 
in the mass region 11 $\leq$ A $\leq$ 75, based on the AME2012 data \cite{AME2012}. The Set I (III) and II (IV) are obtained by fitting the data, respectively, without and with the Coulomb exchange term (including the proton self-interaction correction $Z(Z-1)$). The last column $\sigma$ provides the rms deviations from the fitting procedure.}}
\begin{tabular}{c c c c c c}
\hline
\hline
$     $  &~~~Coulomb energy $E_{c}$ &~~~$a_{c}$ (MeV) &~~~$b$ & $\sigma$ (keV)~~\\
\hline
Set I &~~~$a_{c} {Z^{2} \over A^{1/3}}$      &~~~$0.625$    &~~~$-$~~& 336 \\
\hline
Set II &~~~$a_{c} {Z^{2} \over A^{1/3}}(1 - bZ^{-2/3})$      &~~~$0.715$    &~~~$1.374$~~& 121 \\
\hline
Set III &~~~$a_{c} {Z(Z - 1) \over A^{1/3}}$      &~~~$0.642$    &~~~$-$~~ & 249\\
\hline
Set IV &~~~$a_{c} {Z(Z - 1) \over A^{1/3}}(1 - bZ^{-2/3})$      &~~~$0.704$    &~~~$0.985$~~ &118 \\
\hline
\hline
\end{tabular}
\label{table1}
\end{table}

\subsection{Coulomb energy correction}
\begin{figure}
\centering
\includegraphics[width=0.95\linewidth]{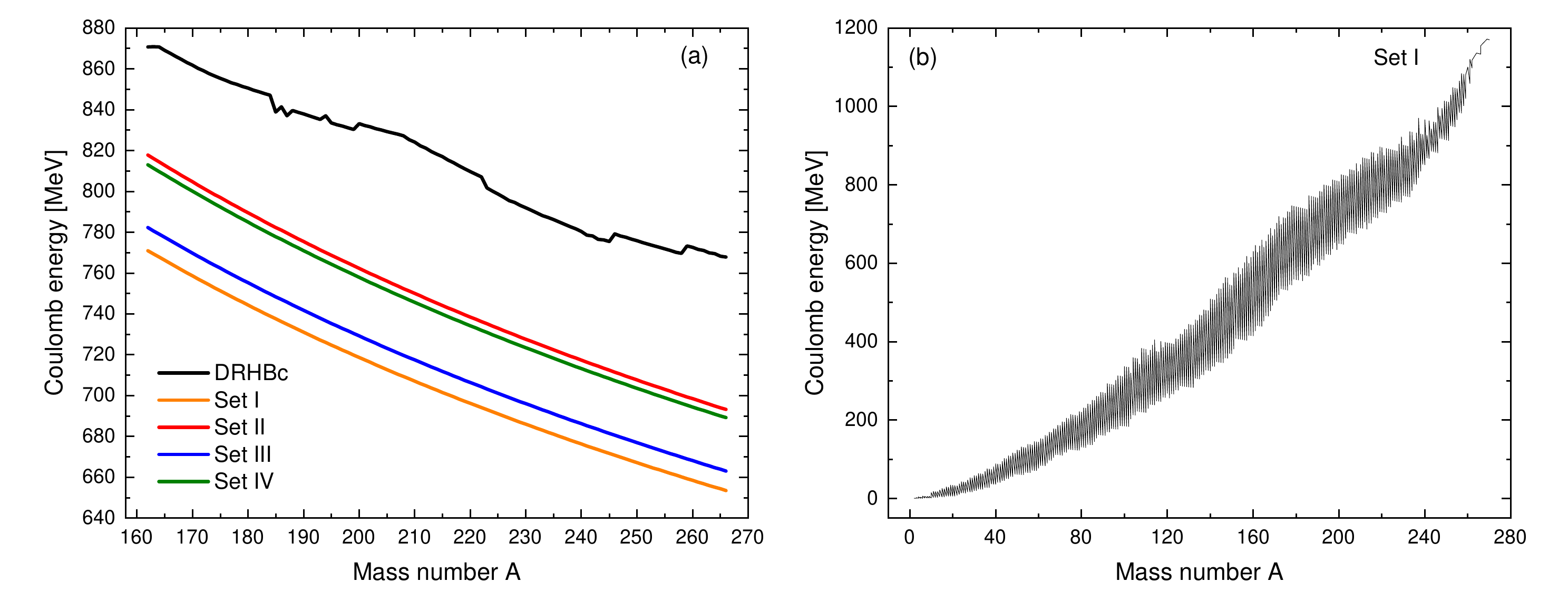}
\caption{(Color online) Coulomb energies of Pb isotopes (a). Set I $\sim $ IV in Table \ref{table1} are taken from Ref. \cite{Tian2014} and the DRHBc Coulomb energies are taken from Ref. \cite{Kaiyuan2022}. The Coulomb energy data in the panel (b) are calculated for all isotopes, for a given mass number $A$, using Set I. The Coulomb energies of all isobars for a given mass number $A$ are indicated explicitly.}
\label{fig4}
\end{figure}
\begin{figure}
\centering
\includegraphics[width=0.95\linewidth]{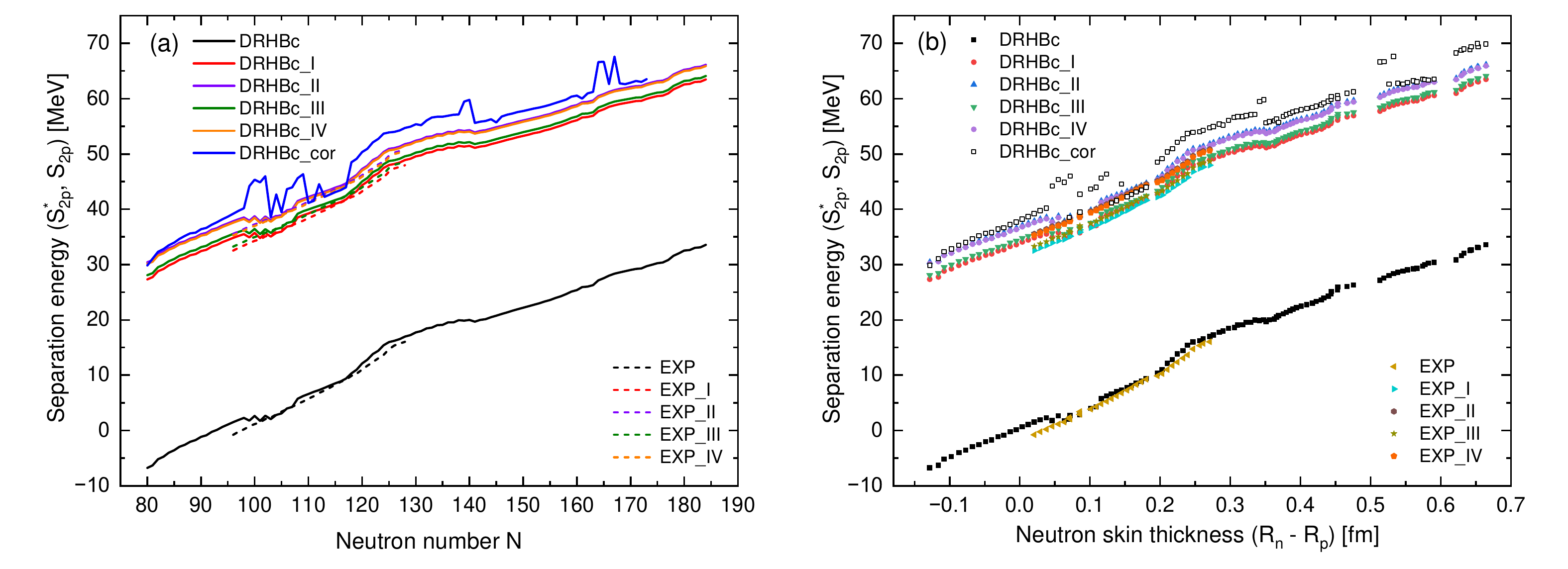}
\caption{(Color online) Evolution of $S_{2p}$ (lower region) and $S_{2p}^{*}$ (upper region) for Pb isotopes versus $N$ (a) and NST (b). DRHBc\_I $\sim$ DRHBc\_IV label the results obtained by subtracting the Coulomb energy from Set I $\sim$ Set IV. DRHBc\_cor is the result obtained by subtracting the Coulomb energy calculated by the DRHBc model itself. EXP [dashed lines in (a)  and yellow symbols in (b)] are the experimental data from AME2020. EXP\_I $\sim$ EXP\_IV are the experimental data (AME2020) corrected by subtracting the Coulomb energy from Set I $\sim$ Set IV.}
\label{fig5}
\end{figure}

Since the Coulomb energy affects the proton separation energy, we subtract the Coulomb energy.  Following the prescriptions used in Ref. \cite{Tian2014}, which are summarized in Table \ref{table1}, we show the evolution of the Coulomb energy of Pb isotopes as well as that of the DRHBc calculation in Fig. \ref{fig4}. The Coulomb energy decreases with the mass number, as expected. We also note that the Set I and III are characterized by a Coulomb energy which is
about 40 $\sim$ 50 MeV smaller than the Set II and IV. This comes from the fact that the exchange Coulomb energy cancels with the direct Coulomb energy, but the $a_c$ values themselves in Set II and IV are larger than those in the Set I and III. Here we note that the isospin symmetry   breaking interactions  are not included  in the results of Fig. \ref{fig4}.

The Coulomb energy by the DRHBc is presented with the black curve, and the values are much larger 
than all what is obtained with the Set I $\sim$ Set IV
in Table \ref{table1}. 
In fact, these include the Coulomb exchange energy, either explicitly (Set II and IV) or implicitly (Set I and III). Therefore,
more refined calculations of the Coulomb exchange term might be desirable for more exact access of the Coulomb energy, because the Coulomb exchange interaction is not included in the DRHBc model. The small discontinuous Coulomb energy by the DRHBc comes from the deformation of Pb isotopes. The Coulomb energy per proton in Pb isotopes, from the DRHBc model, is decreasing from  about 11 to 8 MeV along  the $^{162 - 266}$Pb isotopes. Figure \ref{fig4}(b) shows an example of the Coulomb energy from the Set I in the whole mass table: it increases with the mass number as expected. The spread in each isobaric chain amounts to a few tens of MeV. It means that the Coulomb energy correction is important in the present calculation.

Figure \ref{fig5}(a) shows the $2p$ separation energy ($S_{2p}$) and the Coulomb-corrected  one ($S_{2p}^{*}$).   The latter is  larger, by about 36 MeV on average, than $S_{2p}$. We can find a clear correlation of $S_{2p}$ and  $S_{2p}^{*}$ with the neutron number, that is, even when the Coulomb energy is subtracted. Figure \ref{fig5}(b) provides the $S_{2p}$ and $S_{2p}^{*}$ evolution in Pb isotopes as a function of the NST. The correlation is still evident. The uncertainty from the Coulomb energy estimation is about 5 MeV (see the upper curves in Fig. \ref{fig5}). We also present the experimental separation energy from AME2020 data \cite{AME2020}.  

\begin{figure}
	\centering
	\includegraphics[width=0.48\linewidth]{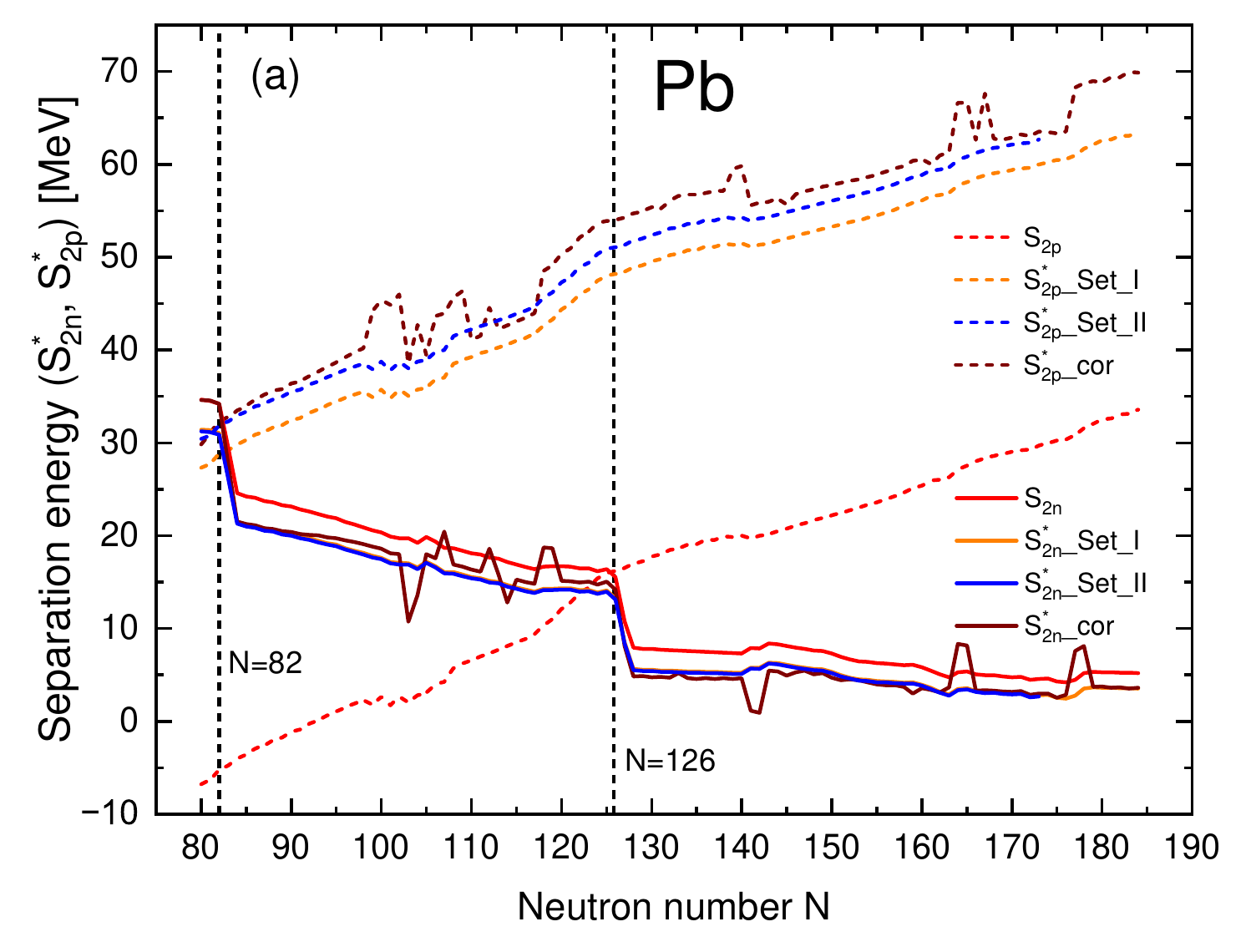}
	\includegraphics[width=0.48\linewidth]{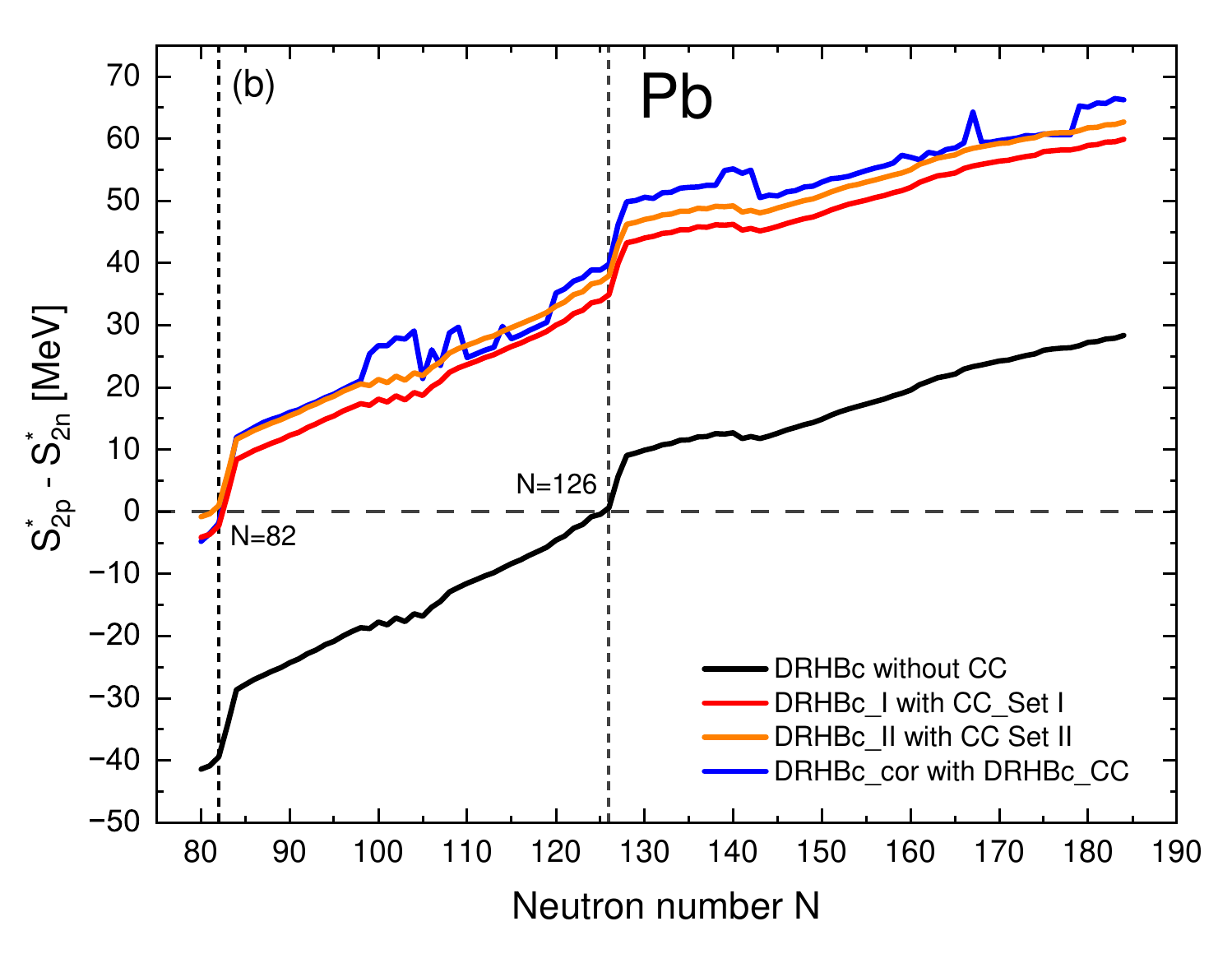}
	\includegraphics[width=0.48\linewidth]{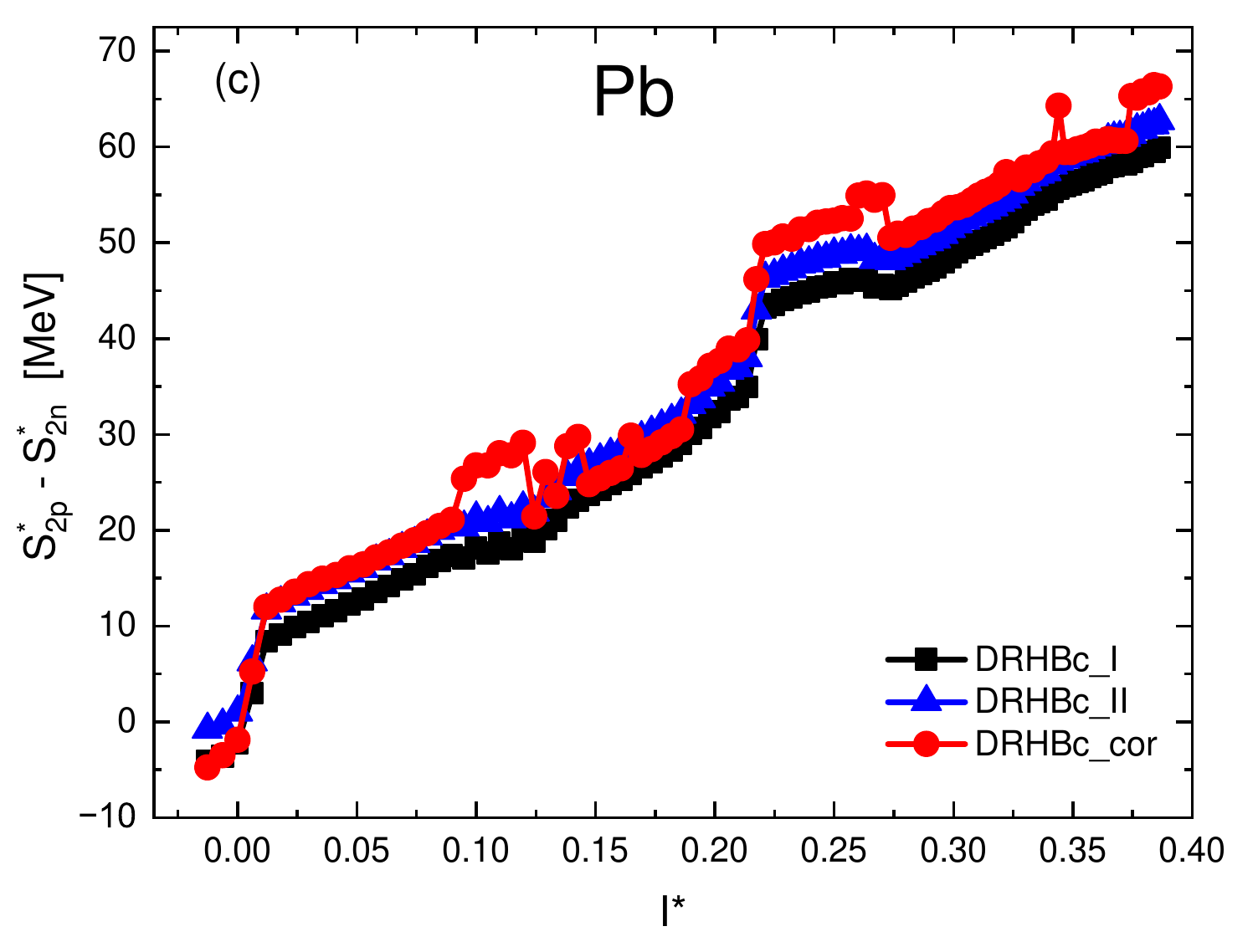}
	\includegraphics[width=0.48\linewidth]{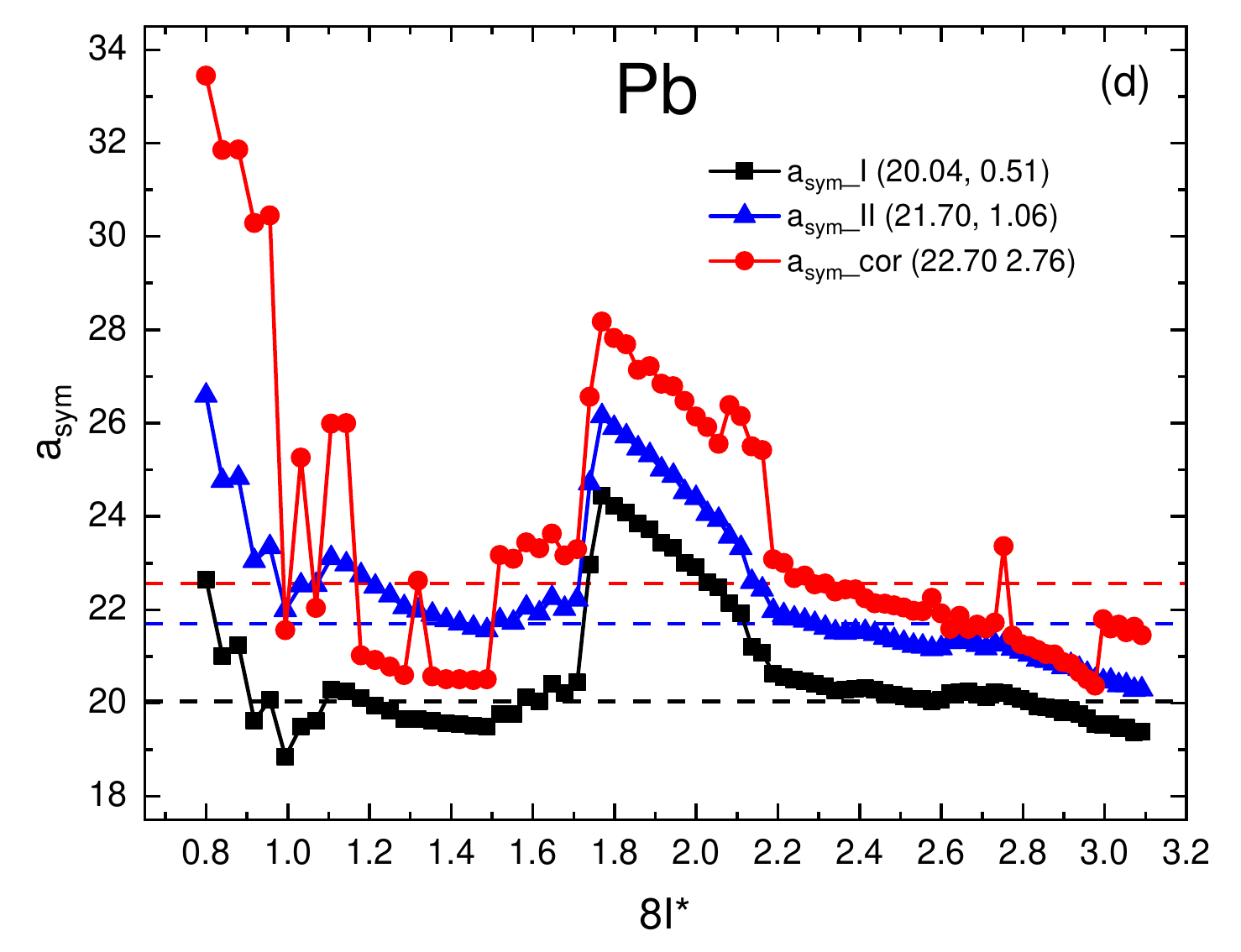}
	\caption{(Color online) (a) Evolutions of  $S_{2p}$, $S_{2p}^{*}$, $S_{2n}$ and $S_{2n}^{*}$  as a function of the neutron number $N$ and (b) the  difference $S_{2p}^{*}$ - $S_{2n}^{*}$ for the Pb isotopes. The values with asterisks are those including the Coulomb energy correction. Panels (c) and (d) show the difference $S_{2p}^{*}$ - $S_{2n}^{*}$ as a function of $I^{*}= (N-Z)/(A-2)$, and $a_{sym}$ w.r.t $8 I^*$, respectively. The numbers in the parenthesis in panel (d) are the central values of $a_{sym}$ and the root mean square (rms) deviations.  Here we show only the results from Set I and II because  their rms deviations $\sigma$ are smaller than those of Set III and IV including the proton self-interaction correction.}
	\label{fig6}
\end{figure}

Figure \ref{fig6} (a) and (b) show the evolution of $S_{2p}$, $S_{2p}^*$,  $S_{2n}$  and $S_{2n}^{*}$ for Pb isotopes, as well as the difference of $S_{2p}^*$ and $S_{2n}^{*}$, as a function of the neutron number $N$. $S_{2p}$ and $S_{2n}$ show a $X$-type crossing behaviour at the  point near the magic shell $N=126$, while $S_{2p}^*$ and $S_{2n}^{*}$ cross at the magic number  $N=82$ in the panel (a). We note that $S_{2n}^{*}$ is also affected 
by about 1 $\sim$ 2 MeV by the Coulomb energy subtraction, because of the change of the total potential energy curve.  

The differences between  $S_{2p}^*$ and $S_{2n}^{*}$ in Fig. \ref{fig6}(b) and (c) are found to be  linearly proportional to the neutron number $N$, and also to $I^{*} = (N-Z)/(A-2)$, except at the shell closure 
and in the deformed region. This difference should be constant with respect to $8 I^{*}$ according to the formula in Eq. (\ref{eq:sseq1}), and the constant is directly related to the asymmetry coefficient $a_{sym}$ in the mass formula. Although we can see  large fluctuations,  we extracted the value $a_{sym}$ from Fig. \ref{fig6}(d) by taking a fit to the data. In the fitting process,  we discarded the region $I^{*} < 0.100$, because it is the proton-rich region, and also the $0.217 < I^* < 0.271$ region because it is the region near the magic number.  The values of $a_{sym}$ from Pb isotopes are shown in Fig. \ref{fig6}(d): these span the interval between
20.0 and 22.7 MeV, depending on how the Coulomb correction is implemented. Detailed values are in Table \ref{table2}.

\begin{figure}
	\centering
	\includegraphics[width=0.48\linewidth]{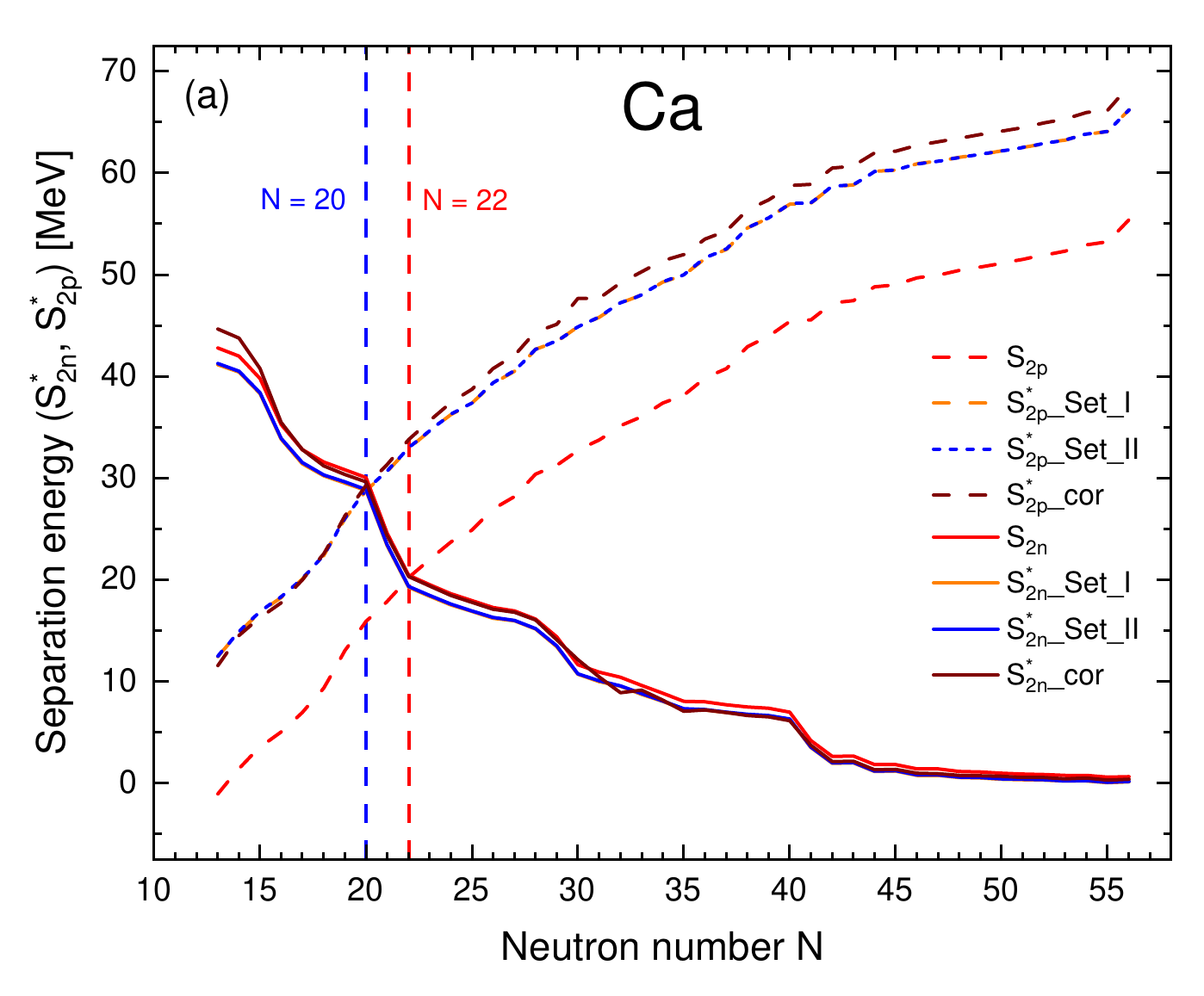}
	\includegraphics[width=0.48\linewidth]{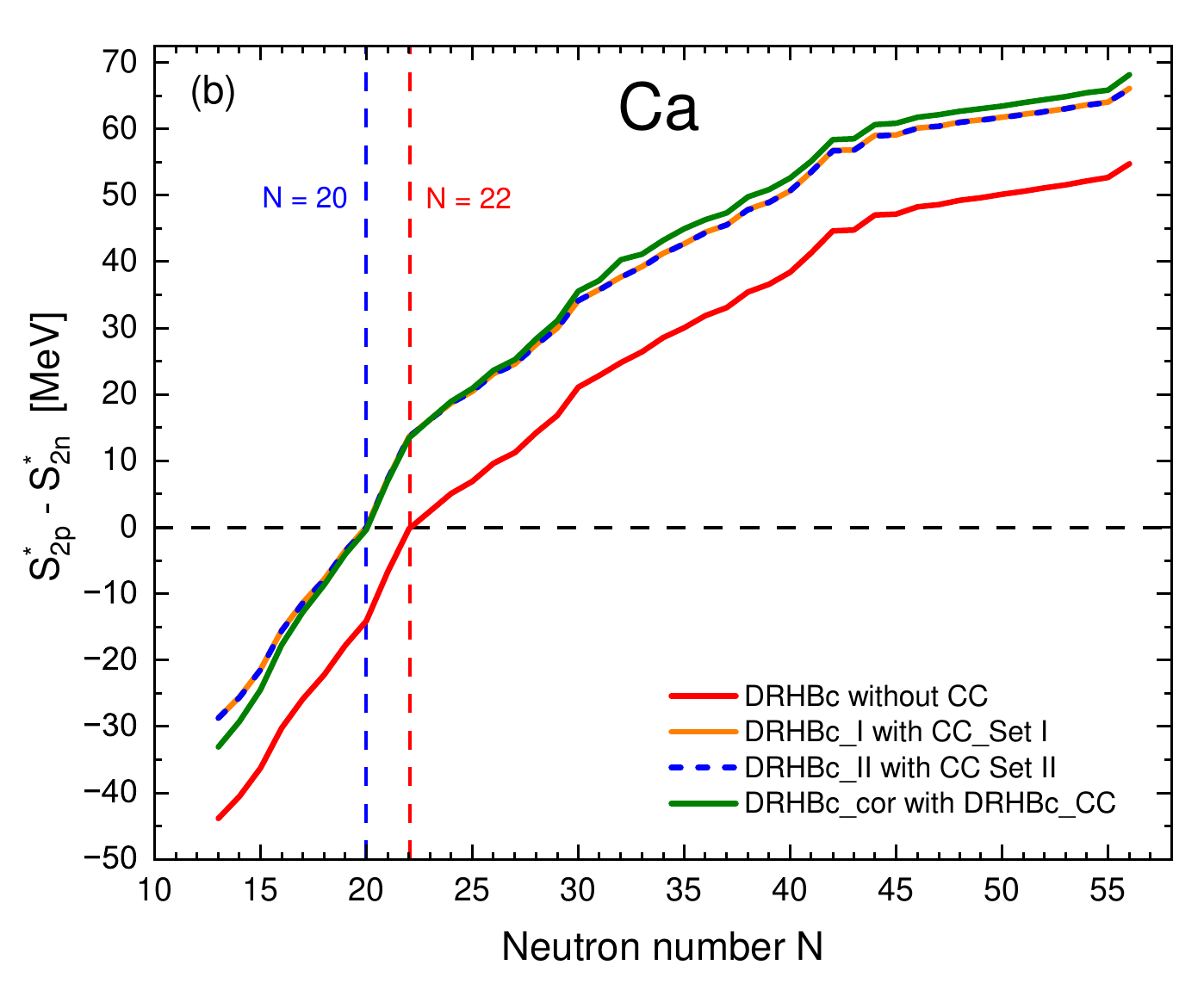}
	\includegraphics[width=0.48\linewidth]{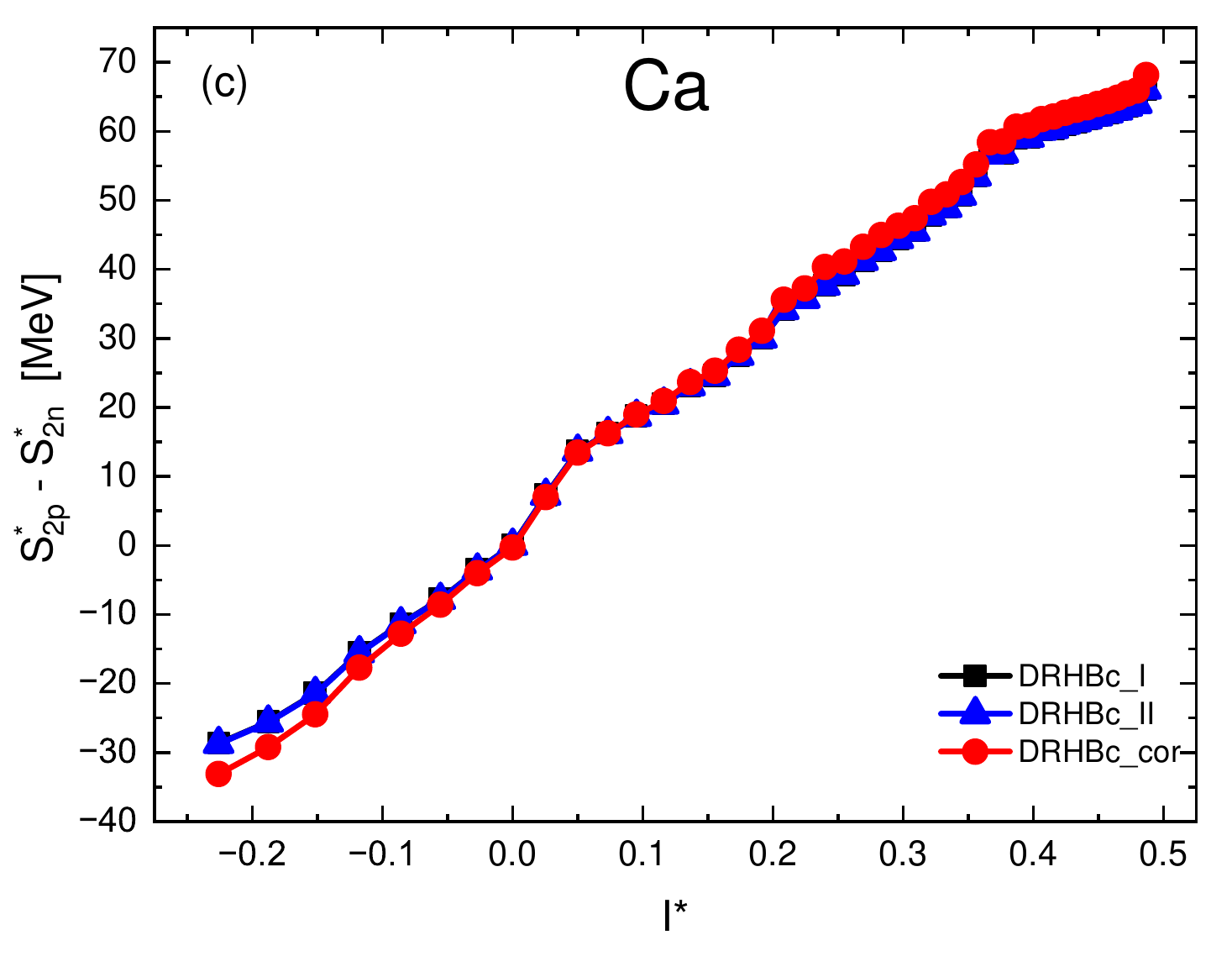}
	\includegraphics[width=0.48\linewidth]{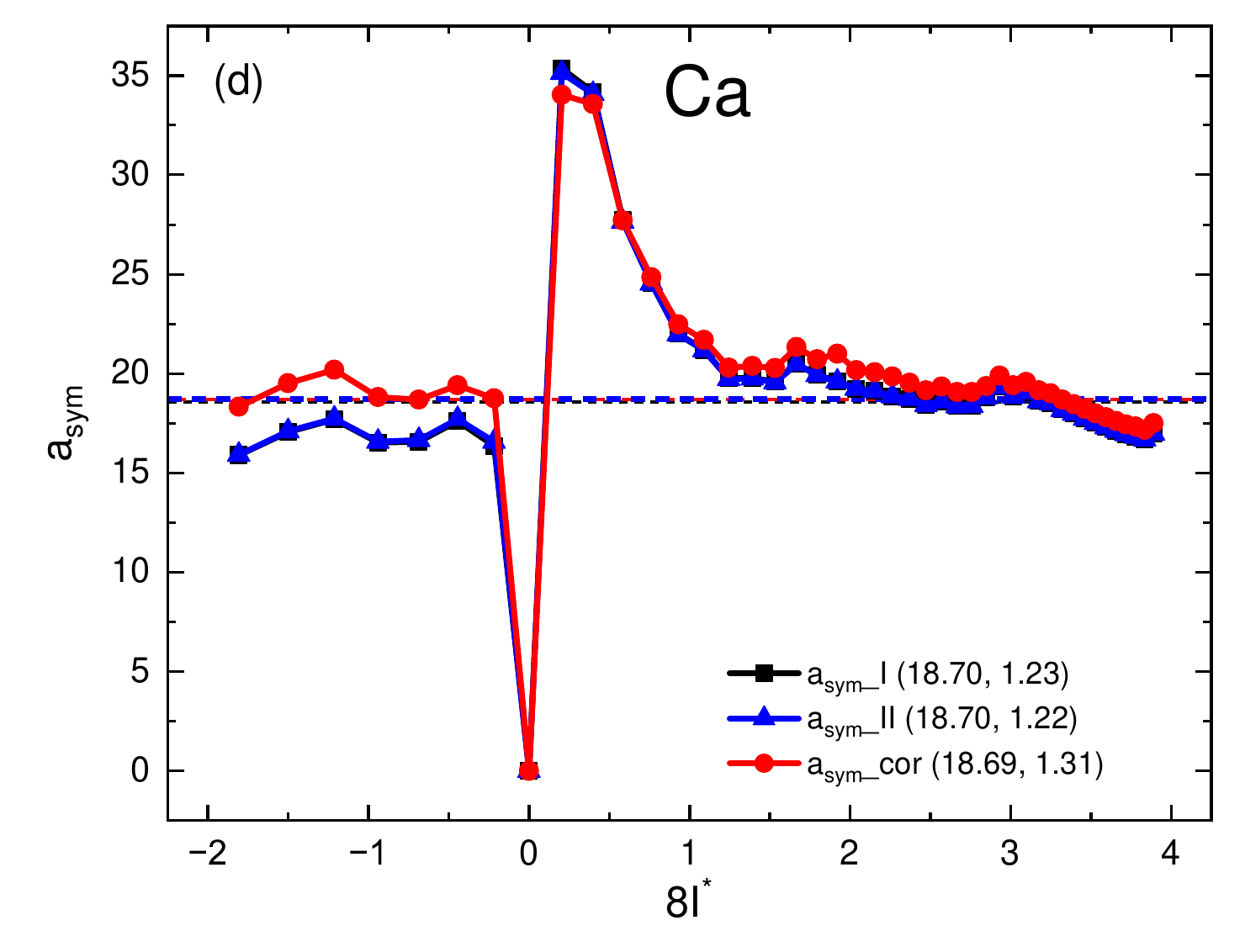}
	\caption{(Color online) Same as Fig. \ref{fig6}, but for Ca isotopes. The numbers in parenthesis are the central values and the rms deviations.}
	\label{fig7}
\end{figure}

In the following, we present the $S_{2p}^{*}$ and $S_{2n}^{*}$ results for Ca isotopes in Fig. \ref{fig7}. The general behaviour is very similar to Pb isotopes. We found also the $X$-type band of $S_{2p}$ and  $S_{2n}$ as well as $S_{2p}^*$ and  $S_{2n}^{*}$  in Fig. \ref{fig7}(a). An interesting point is the change of the crossing point without and with the Coulomb energy. With the Coulomb energy subtraction, the crossing point moves to the magic number $N=20$ for $S_{2p}^*$ from $N=22$ for $S_{2p}$. This is very similar to the results for Pb isotopes in the vicinity of the magic number $N=82$ and 126. This tendency implies that both separation energies, $S_{2p}^*$ and  $S_{2n}^{*}$, are the same for $N=Z$ nuclei, and that the present approach is very reasonable. 

However, the crossing points of $S_{2p}$ and $S_{2n}$ (or $S_{2n}^*$) are at $N= $126 and $N=$ 22, respectively, for Pb and Ca isotopes. 
 Other results from Set III and IV show similar features. Even the results of AME and FRDM model do not provide different outcome. Since $N=$ 126 for $^{208}$Pb is the magic number, the value $N=$ 22 is a bit difficult to understand. 

The relative difference $S_{2p}^{*}$ -- $S_{2n}^{*}$ in Ca isotopes, in Fig. \ref{fig7}(b) and (c), shows also a clear correlation with the neutron number and $I^*$. The $a_{sym}$ from Fig.\ref{fig7}(d) is estimated as about 18.7 $\sim$ 19.7 MeV, and these values are a bit smaller than those from the Pb isotopes.
In the fitting we also discarded the region around $I^* = 0$, corresponding to the $N=Z=20$ case. The final $a_{sym}$ values fitted  from Fig. \ref{fig6}(d) and Fig. \ref{fig7}(d) are also tabulated in Table \ref{table2} with the results of other mass tables, AME2020 and FRDM2012. In particular, we note that the DRHBc\_cor value is larger, by about 1 MeV, than those from the other Coulomb corrected Set I and II.        

\begin{table}[h]
	\caption{Results of  extracted $a_{sym}$ for Ca and Pb with DRHBc, AME2020 and FRDM2012. The numbers in parenthesis are the rms deviations.}
	\begin{tabular}{c c c c}
		\hline
		\hline
		&~~~DRHBc &~~~AME2020 &~~~FRDM2012~~\\
		\hline
		Ca\_$a_{sym}$\_I &~~18.70~(1.23)&~~~$~~18.96~(0.73)$  &~~~$~~19.66~(0.48)$~~ \\
		\hline
		Ca\_$a_{sym}$\_II &~~18.69~(1.22)&~~~$~~18.93~(0.72)$    &~~~$~~19.65~(0.47)$~~ \\
		\hline
		Ca\_$a_{sym}$\_cor &~~19.34~(1.31)&~~~$~~-$     &~~~$~~-$~~ \\
		\hline
		Pb\_$a_{sym}$\_I &~~20.04~(0.51)&~~~$~~19.60~(0.41)$     &~~~$~~20.67~(0.76)$~~ \\
		\hline
		Pb\_$a_{sym}$\_II &~~21.70~(1.06)&~~~$~~22.15~(0.68)$    &~~~$~~22.33~(0.41)$~~ \\
		\hline
		Pb\_$a_{sym}$\_cor &~~22.70~(2.76)&~~~$~~-$   &~~~$~~-$~~ \\
		\hline												
		\hline
	\end{tabular}
	\label{table2}
\end{table}

\subsection{Surface symmetry energy correction}


%
Hereafter, we will disentangle the contribution of the surface symmetry energy to $a_{sym}$ by comparing the two results of Pb and Ca isotopes. We will try to extract a model-independent $a_{sym}^v$ value. We use the $S_{2p}^{*}$ -- $S_{2n}^{*}$ formula given by Eq. (\ref{eq:sseq2}), which divides the symmetry energy into volume ($a_{sym}^{v}$) and surface ($a_{sym}^{s}$) part. By using the ratio of $a_s /a_v$ from nuclear models \cite{Lunn2003}, we deduce the $a_{sym}^v$ values and tabulate them in Table \ref{table3}.

If we compare the results in Table \ref{table2} to those in \ref{table3} (see the first 6 rows for the DRHBc case), the $a_{sym}^v$ value is increased compared to the $a_{sym}$ value, by about 7.5 MeV for Ca and 4.5 MeV for Pb isotopes, due to the surface symmetry energy ($a_{sym}^{s}$) correction, in the DRHBc model case. If we use a bit larger value of $a_s / a_v$, from FRDM, the change amounts to a larger value, 10.86 and 6.62 for Ca and Pb isotopes. It is interesting that the corrections for Ca isotopes are larger than those by Pb isotopes. {It suggests  that the surface contribution to the symmetry energy becomes larger in light nuclei rather than those by heavy nuclei as expected from Eq. (5)}.

\subsection{Results from other mass models}

\begin{figure}
	\centering
	\includegraphics[width=0.48\linewidth]{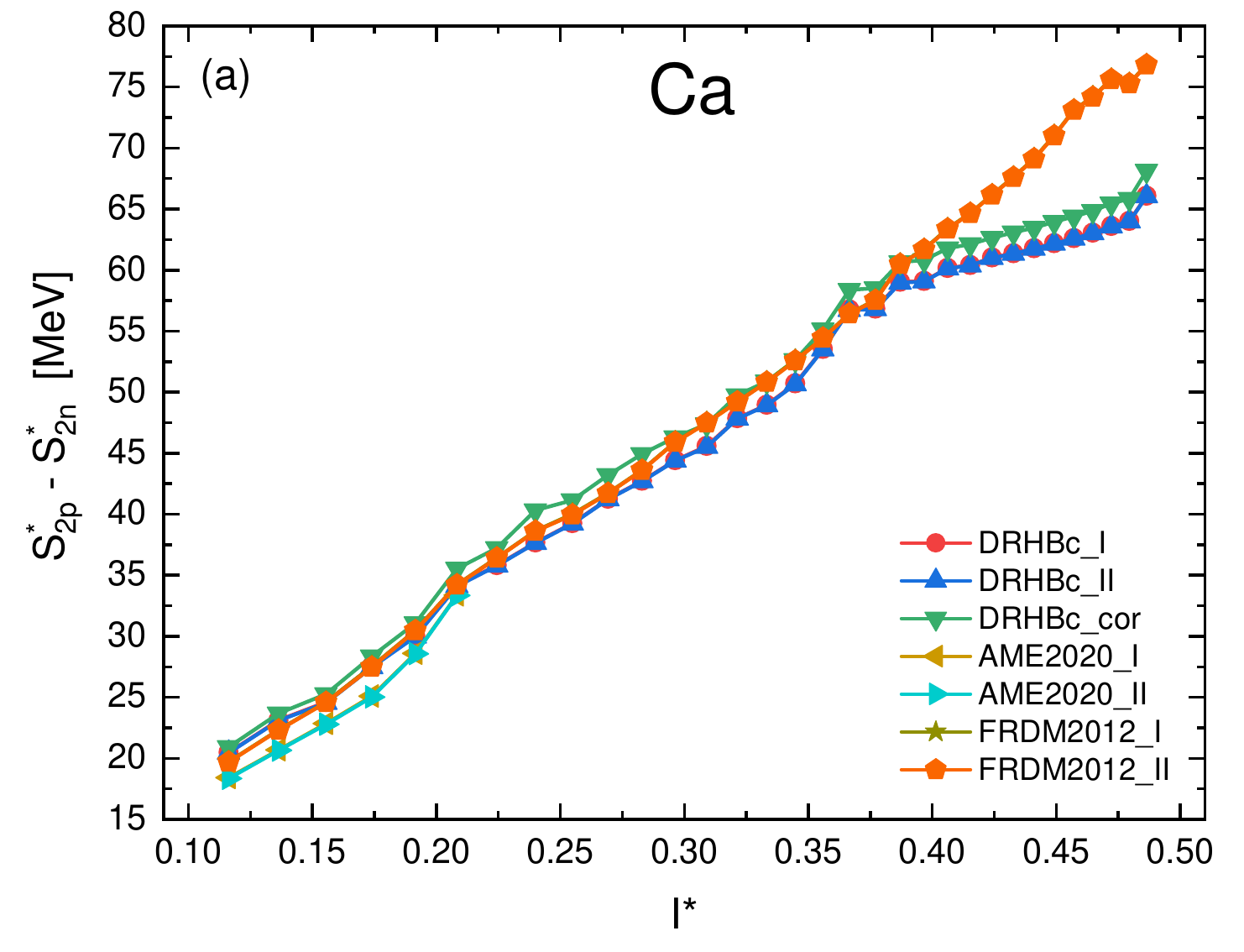}
	\includegraphics[width=0.48\linewidth]{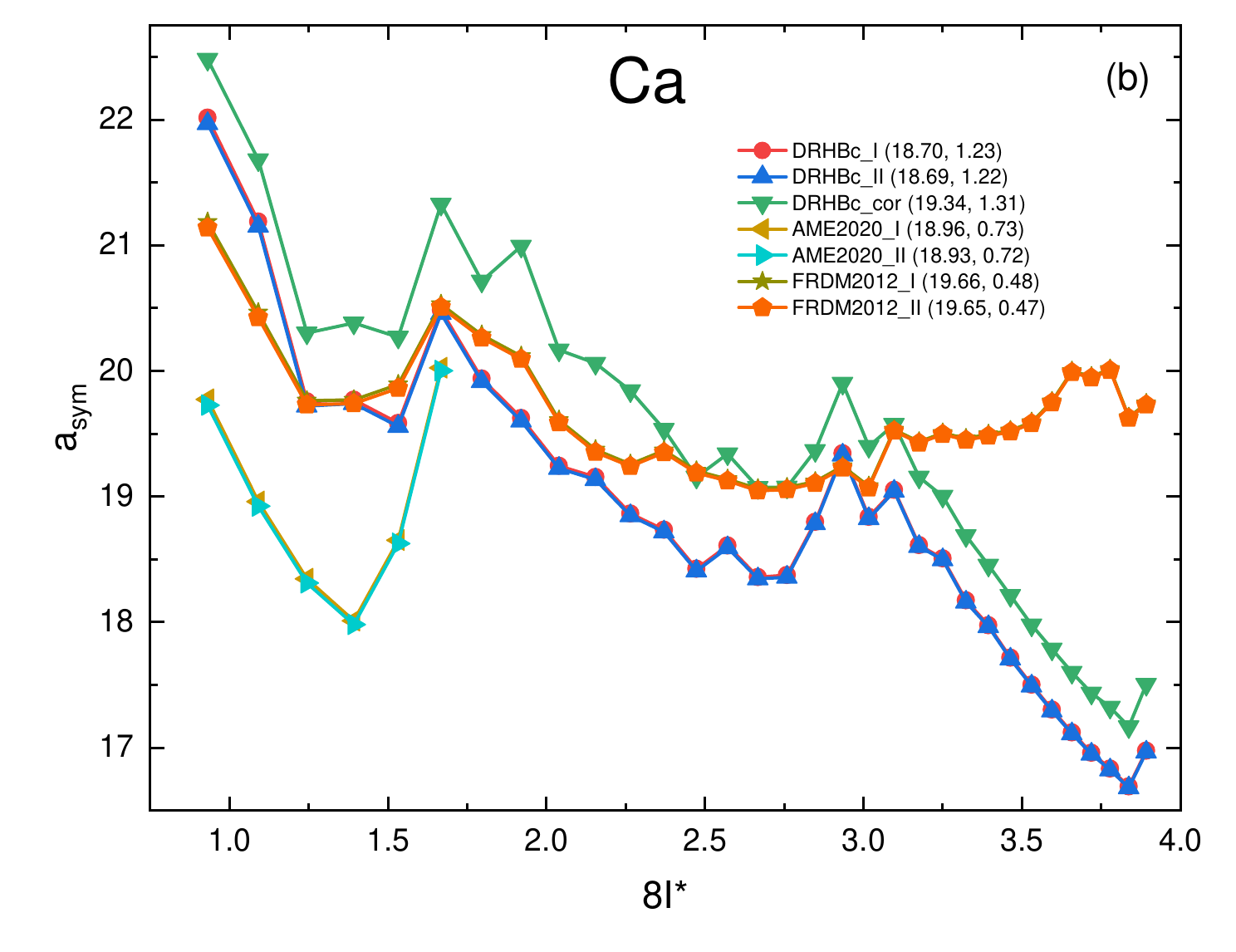}
	\includegraphics[width=0.48\linewidth]{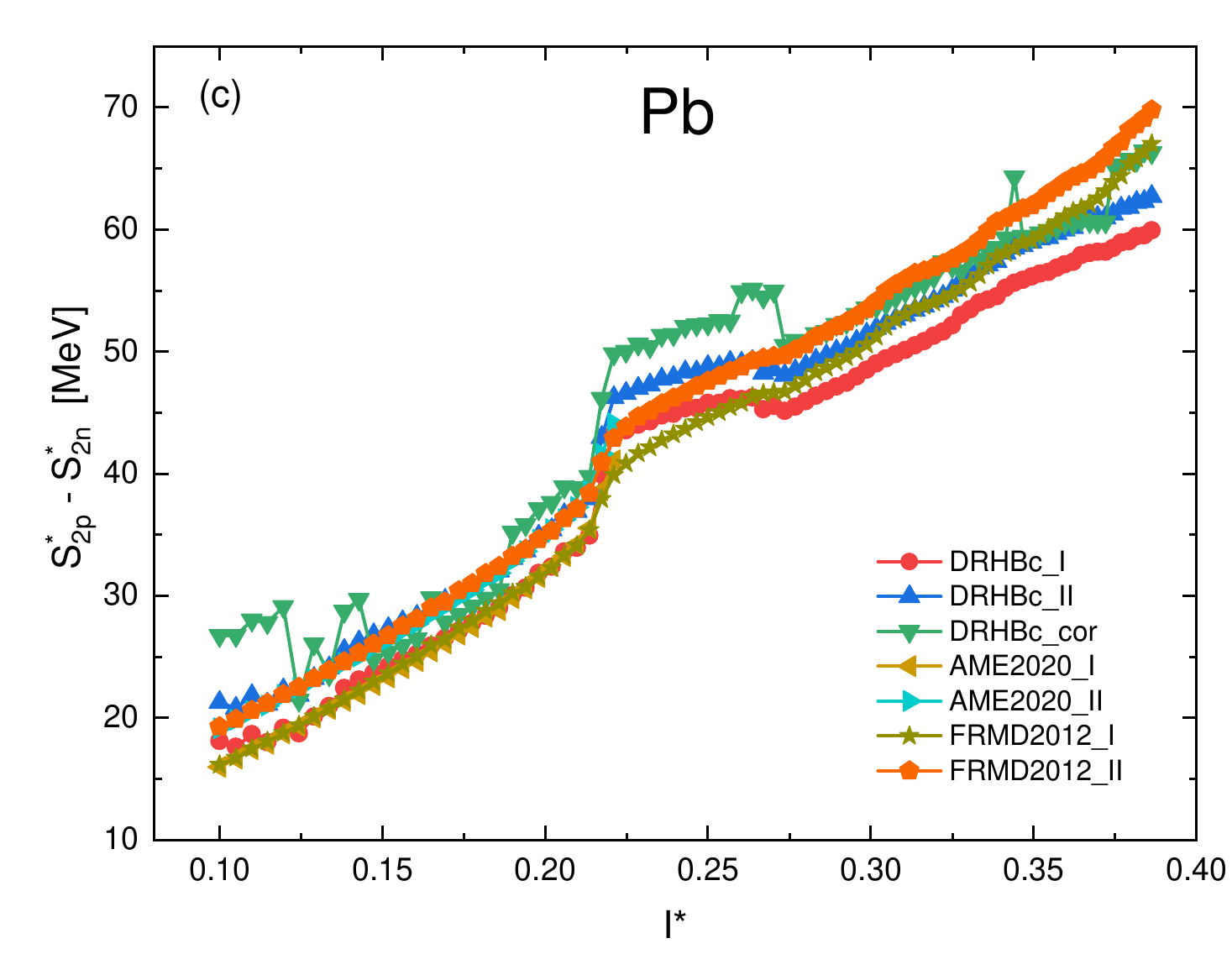}
	\includegraphics[width=0.48\linewidth]{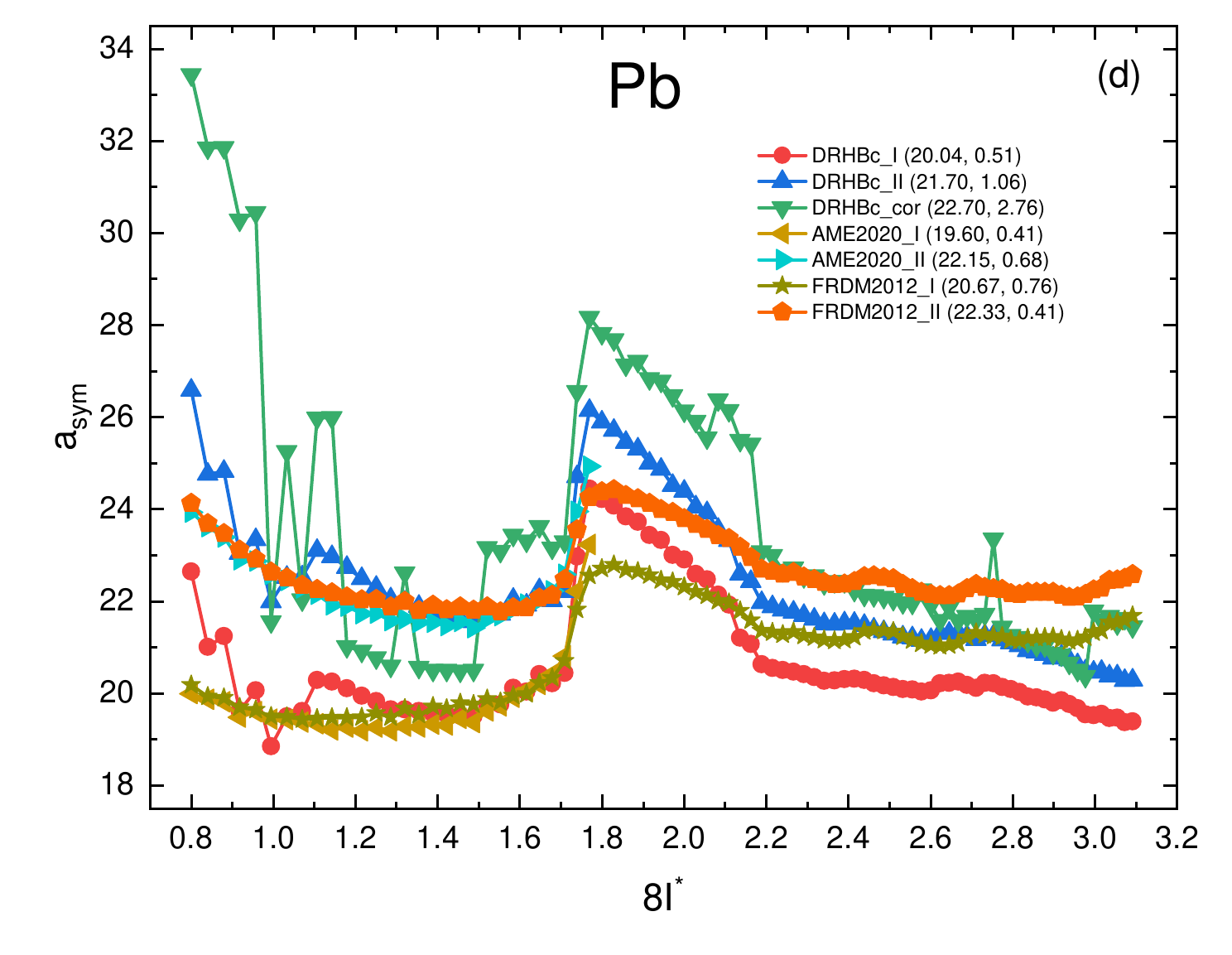}
	\caption{(Color online) $S_{2p}^* - S_{2n}^*$ (a) and $a_{sym}$ (b) for Ca from DRHBc, AME2020 and FRDM2012, which correspond to the results in Fig. \ref{fig7} (c) and (d). $S_{2p}^* - S_{2n}^*$ (c) and $a_{sym}$ (d) for $^{208}$Pb from DRHBc, AME2020 and FRDM2012, which correspond to the results in Fig. \ref{fig6} (c) and (d). The numbers in the parenthesis of panels (b) and (d) are the central values and the rms deviations.}
\label{fig8}
\end{figure}

\begin{table}[h]
\caption{Macroscopic parameters ($a_{s} / a_{v}$) deduced from different mass formulas \cite{Lunn2003}, and the results of $a_{sym}^{v}$ extracted after correcting $a_{sym}$ by the surface contribution.
We start from $a_{sym}$ values for Ca and Pb isotopes, as tabulated in Table \ref{table2}, coming from the DRHBc, AME2020 and FRDM2012 approaches. The numbers in the parenthesis are rms deviations.}

\begin{tabular}{c c c c }
		\hline
		\hline
		 &~~~HFB   &~~~FRDM   &~~~Ref. \cite{Lunn2003}~~\\
		\hline
	$a_{s}/a_{v}$      &~~~$1.11$        &~~~$1.41$      &~~~$1.13$~~ \\
		\hline\hline
	DRHBc & & & \\	
		\hline\hline
		Ca\_$a_{sym}^{v}$\_I      &~~~$~~26.27~(2.28)$      &~~~$~~29.56~(2.83)$     &~~~$~~26.49~(2.31)$~~ \\
		\hline
		Ca\_$a_{sym}^{v}$\_II      &~~~$~~26.25~(2.26)$       &~~~$~~29.53~(2.82)$     &~~~$~~26.47~(2.30)$~~ \\
		\hline
		Ca\_$a_{sym}^{v}$\_cor      &~~~$~~27.17~(2.40)$      &~~~$~~30.57~(2.30)$     &~~~$~~27.40~(2.44)$~~ \\
		\hline		
		Pb\_$a_{sym}^{v}$\_I     &~~~$~~24.53~(0.71)$       &~~~$~~26.14~(0.80)$     &~~~$~~24.64~(0.71)$~~ \\
		\hline
		Pb\_$a_{sym}^{v}$\_II      &~~~$~~26.57~(1.52)$      &~~~$~~28.31~(1.70)$     &~~~$~~26.69~(1.53)$~~ \\
		\hline
		Pb\_$a_{sym}^{v}$\_cor      &~~~$~~27.80~(3.56)$      &~~~$~~29.63~(3.87)$     &~~~$~~27.93~(3.58)$~~ \\
		\hline															
		\hline
	AME2020 & & & \\			
		\hline
		\hline
		Ca\_$a_{sym}^{v}$\_I     &~~~$~~27.49~(1.07)$       &~~~$~~31.36~(1.23)$     &~~~$~~27.74~(1.08)$~~ \\
		\hline
		Ca\_$a_{sym}^{v}$\_II     &~~~$~~27.44~(1.06)$      &~~~$~~31.30~(1.23)$     &~~~$~~27.70~(1.07)$~~ \\
		\hline
		Pb\_$a_{sym}^{v}$\_I      &~~~$~~24.24~(0.48)$      &~~~$~~25.93~(0.50)$     &~~~$~~24.36~(0.48)$~~ \\
		\hline
		Pb\_$a_{sym}^{v}$\_II    &~~~$~~27.40~(0.90)$     &~~~$~~29.31~(0.99)$     &~~~$~~27.54~(0.91)$~~ \\
	\hline\hline
	FRDM2012 & & & \\			
		\hline
		\hline
Ca\_$a_{sym}^{v}$\_I      &~~~$~~27.59~(1.08)$       &~~~$~~31.03~(1.47)$     &~~~$~~27.82~(1.10)$~~ \\
		\hline
		Ca\_$a_{sym}^{v}$\_II      &~~~$~~27.57~(1.06)$      &~~~$~~31.00~(1.46)$     &~~~$~~27.79~(1.09)$~~ \\
		\hline
		Pb\_$a_{sym}^{v}$\_I     &~~~$~~25.30~(0.72)$     &~~~$~~26.95~(0.61)$     &~~~$~~25.41~(0.72)$~~ \\
		\hline
		Pb\_$a_{sym}^{v}$\_II     &~~~$~~27.34~(0.61)$     &~~~$~~29.13~(0.71)$     &~~~$~~27.46~(0.61)$~~ \\
		\hline													
		\hline

\end{tabular}
\label{table3}
\end{table}

Here we perform the evaluation of $S_{2p}^* - S_{2n}^*$ and $a_{sym}$ for Ca and Pb by using each of the nuclear mass models, AME2020 and FRDM2012. For example, Fig. \ref{fig8} presents the results of $S_{2p}^* - S_{2n}^*$ and $a_{sym}$ from other mass models. The final results for $a_{sym}^{v}$, corrected by the surface contribution,  are  tabulated in Table \ref{table3} for the nuclear mass models. The results for $a_{sym}^{v}$ depend on the $a_s / a_v$ value. {All of the results for $a_{sym}^v$ are summarized in Table \ref{table3}. For example, the values from  FRDM2012 are larger than the values from other models. Final results of $a_{sym}$ and $a_{sym}^v$ with the ratio $a_s/a_v$ extracted from the mass tables are tabulated and compared to other results \cite{Kim2023,Tian2014} in Table \ref{table4}.}

\begin{table}[h]
	\caption{Summary of $a_{sym}$, $a_{sym}^v$ and $a_s / a_v$ for Ca and Pb with DRHBc, AME2020 and FRDM2012. Kim I and Kim II \cite{Kim2023} are the results by using full range mass data, respectively, by DRHBc and AME2020 data. Tian \cite{Tian2014} are taken by AME2012 data \cite{AME2012}.}
	\begin{tabular}{c c c c}
		\hline
		\hline
		&~~~$a_{sym}$ &~~~$a_{sym}^v$ &~~~$a_s/a_v$~~\\
		\hline\hline
		Ca~(DRHBc) &~~18.7 $\sim$\footnotemark[1] 19.3&~~~ 26.2 $\sim$ 27.4 ($\sim$ 30.6)\footnotemark[2] &~~~1.10 $\sim$ 1.13~~($\sim$ 1.41)\footnotemark[2] \\
				\hline
		Ca~(AME2020) &~~19.0&~~~ 27.4 $\sim$ 27.7 ($\sim$ 31.4)&~~~1.10 $\sim$ 1.13~~($\sim$ 1.41) \\
		\hline
		Ca~(FRDM2012) &~~19.6 $\sim$ 19.7&~~~ 27.5 $\sim$ 27.8 ($\sim$ 31.3)&~~~1.10 $\sim$ 1.13~~($\sim$ 1.41) \\
		\hline\hline
		Pb~(DRHBc) &~~20.0 $\sim$ 22.7&~~~24.5 $\sim$ 27.9 ($\sim$ 29.6) &~~~1.10 $\sim$ 1.13 ($\sim$ 1.41) \\
		\hline
    	Pb~(AME2020) &~~19.6 $\sim$ 22.1&~~~24.2 $\sim$ 27.5 ($\sim$ 29.3) &~~~1.10 $\sim$ 1.13 ($\sim$ 1.41) \\
		\hline
		Pb~(FRDM2012) &~~20.7 $\sim$ 22.3&~~~25.3 $\sim$ 27.4 ($\sim$ 29.1)&~~~1.10 $\sim$ 1.13 ($\sim$ 1.41) \\
		\hline\hline
		Kim I~(A = 208)\footnotemark[3] \cite{Kim2023} &~~21.36&~~~$27.85$     &~~~$~~1.38$\footnotemark[4]~~ \\
		\hline
		Kim II~(A = 208) \cite{Kim2023} &~~22.32&~~~$28.54$     &~~~$~~1.29$\footnotemark[4]~~ \\
		\hline
		Tian \cite{Tian2014} &~~22.25&~~~~28.32 &~~~~$1.27$\footnotemark[4]~~ \\
		\hline
		\hline
	\end{tabular}
	 \footnotetext[1]{~ The $\sim$ comes from the Coulomb correction recipes in Table IV.}
	  \footnotetext[2]{~ The number in the parenthesis is the result by the FRDM model, which shows a large $a_s / a_v$ = 1.41.}
	  \footnotetext[3]{~Kim I, Kim II and Tian employed a formula $a_{sym} (A) = a_{sym} ( 1 - \kappa A^{1/3})$ with $\kappa = a_{sym}^s/ a_{sym}^v$.} 
	  \footnotetext[4]{~This value is $\kappa = a_{sym}^s / a_{sym}^v$.}
	\label{table4}
\end{table}

\subsection{Model independent approach}

\begin{figure}
	\centering
	\includegraphics[width=0.48\linewidth]{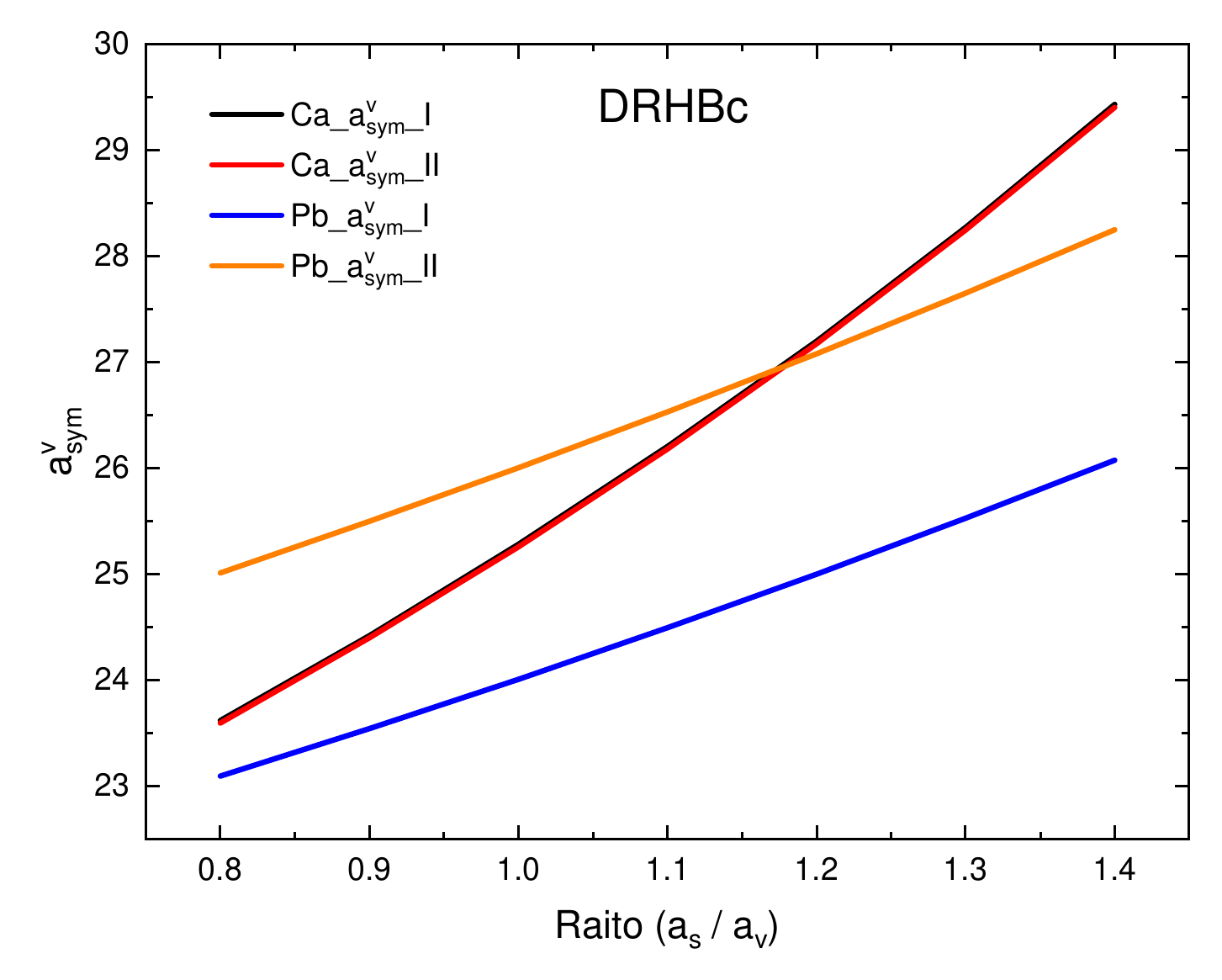}
	\includegraphics[width=0.48\linewidth]{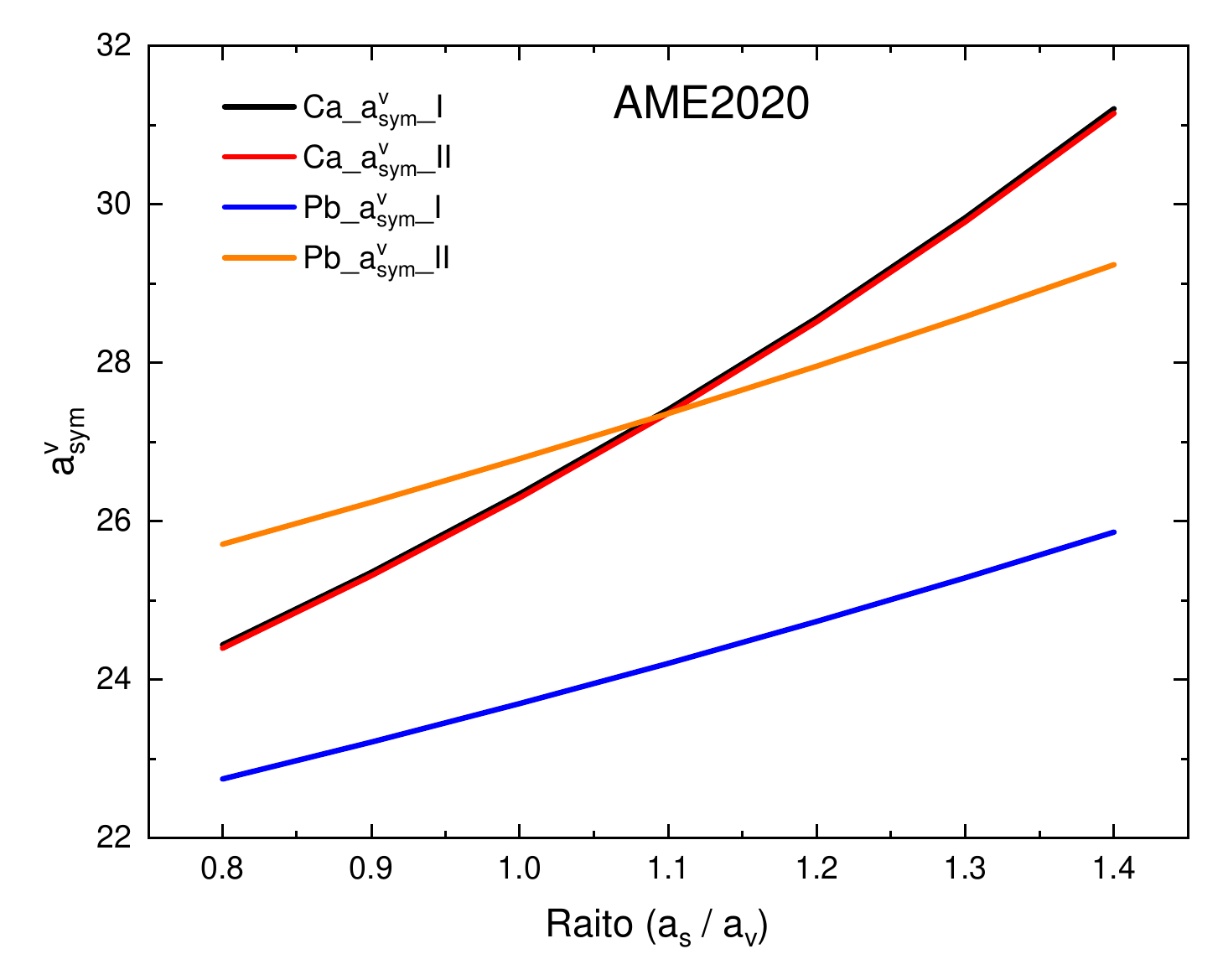}
	\includegraphics[width=0.48\linewidth]{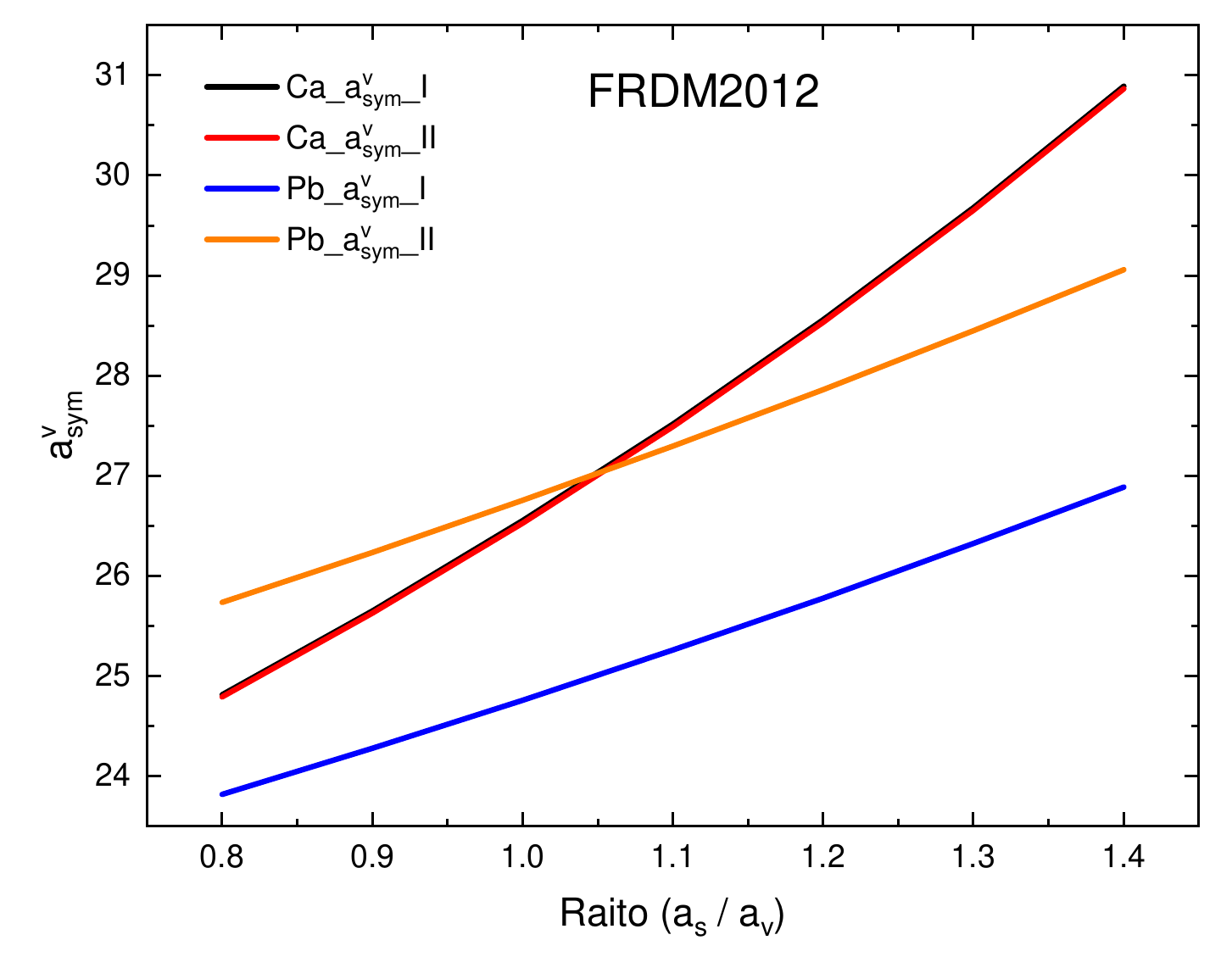}
	\caption{(Color online) $a_{sym}^v$ as a function of the ratio $a_s / a_v$ for each nuclear mass model}
\label{fig9}
\end{figure}

Since the ratio $a_s / a_v$ in Eq. (\ref{eq:sseq2}) is not well determined by other experimental observables, we take it as a free parameter and try to fix $a_{sym}^v$ from the present $a_{sym}$  in Pb and Ca isotopes. The final results are presented in Fig. \ref{fig9}.

Since the Coulomb corrections were performed mainly for light and medium nuclei due to the paucity of mirror nuclei in heavy mass region, we consider only the results for Ca isotopes by Set I and II, and those for Pb isotopes only by Set II (see the large rms deviation in the results by Set I in Table \ref{table1}).
We note that the results from Ca isotopes using Set I and II,
as well as those from Pb isotopes using Set II clearly show a crossing point at $a_{sym}^v =$27 MeV with $a_s / a_v = 1.1 \pm 0.1$. This result is almost model-independent. The results from Pb isotopes using Set I deviate somehow: we should stress, however, that set is characterized by a larger rms deviation (cf. Table \ref{table1}). In addition, the Coulomb corrections may be more meaningful for light and medium-mass nuclei, due
to the paucity of mirror nuclei in the heavy mass region. These issues deserve further investigation.
 
\section{SUMMARY AND CONCLUSION}

Our motivation, in the current study, was to highlight a new, alternative way of deducing the volume and surface symmetry energy coefficients of the nuclear mass formula. Despite many attempts to extract
the symmetry energy from nuclear structure or reaction measurements, or from neutron star observation, the density dependence of the symmetry energy is still plagued by significant uncertainties.
Symmetry coefficients of the mass formula can provide a useful, complementary information. We have extracted them from the proton and neutron separation energies of Pb and Ca isotopes.

First, we have deduced the symmetry energy coefficient,  $a_{sym}$. We have used the mass table provided by the DRHBc model. Our approach could be applied starting from other mass tables.
We have used, then, the ratio $a_s / a_v$ as an input to disentangle the volume and surface symmetry coefficients. 
Our final results for $a_{sym}$ and $a_{sym}^v$ values for Ca and Pb isotopes are tabulated in Table \ref{table4}. 
The $a_{sym}$ values in the present work, obtained from the two-nucleon separation energies, are consistent with other results \cite{Kim2023,Tian2014}.
If we take the central value of $a_{sym}^v$ results in Table \ref{table4}, 
when $a_s / a_v = 1.10 \sim 1.13$, we obtain $a_{sym}^v$ as $a_{sym}^v = 27.0^{+0.96}_{-2.8}$ considering the Pb isotopes, and $a_{sym}^v = 27.0^{+0.8}_{-0.8}$  considering the Ca isotopes. The central value of $a_{sym}^v$ turns out to be independent of the nuclear species; the larger  $a_s / a_v = 1.41$ from FRDM is discarded here.

Since the ratio $a_s / a_v$ is neither determined by nuclear theory, nor by experimental data, we have investigated $a_{sym}^v$ by using the ratio $a_s / a_v$ as a free parameter. Finally, we have obtained $a_{sym}^v = $ 27.0 MeV, almost irrespective of nuclear model and isotopic chain. The ratio  $a_s / a_v$ is constrained also to be $a_s / a_v = 1.10 \sim 1.13$.
From the deduction of the symmetry energy coefficients, $a_{sym}$ and $a_{sym}^{v}$, we may discuss the symmetry energy as well as its slope parameter. We leave it as a future work.

\section{Acknowledgments}
Helpful discussions with members of the DRHBc Mass Table Collaboration are gratefully appreciated. This work was supported by the National Research Foundation of Korea (Grant Nos. NRF-2018R1D1A1B05048026, NRF-2020R1A2C3006177, NRF-2021R1F1A1060066 and NRF-2021R1A6A1A03043957). This work was supported by the National Supercomputing Center with supercomputing resources including technical support (KSC-2022-CRE-0333).

\end{document}